\documentclass[12pt]{article}

\usepackage{graphicx}

\usepackage{amsfonts,amsmath,xspace,amsmath,varioref} %

\usepackage[%
  colorlinks=false %
, pdfpagemode=None %
, hypertex %
]{hyperref} %

\newcommand\llangle{\langle\!\langle}
\newcommand\rrangle{\rangle\!\rangle}

\def\hybrid{\topmargin 0pt      \oddsidemargin 0pt
           \headheight 0pt \headsep 0pt
       \textwidth 6.5in        
       \textheight 9in         
           \marginparwidth 0.0in
           \parskip 5pt plus 1pt   \jot = 1.5ex}
\catcode`\@=11

\newcount\hour
\newcount\minute
\newtoks\amorpm
\hour=\time\divide\hour by60
\minute=\time{\multiply\hour by60 \global\advance\minute by-\hour}
\edef\standardtime{{\ifnum\hour<12 \global\amorpm={am}%
           \else\global\amorpm={pm}\advance\hour by-12 \fi
           \ifnum\hour=0 \hour=12 \fi
           \number\hour:\ifnum\minute<10 0\fi\number\minute\the\amorpm}}
\edef\militarytime{\number\hour:\ifnum\minute<10 0\fi\number\minute}

\def\draft{
  \oddsidemargin -.5truein \def\@oddfoot{{\scriptsize \it \underline{File
        \jobname.tex}. $Revision: 1.7 $} \hfil -- \footnotesize \thepage\ --
    \hfil{\scriptsize\it\today\quad\militarytime}}
  \let\@evenfoot\@oddfoot %
  \overfullrule 3pt %
  \let\label=\draftlabel %
  \let\marginnote=\draftmarginnote %
  \def\@eqnnum{(\theequation)%
    \rlap{\kern\marginparsep\scriptsize\bfseries\@eqnlabel}%
    \global\let\@eqnlabel\@vacuum}
}

\def\numberbysection{\@addtoreset{equation}{section}
           \def\theequation{\thesection.\arabic{equation}}}
\catcode`@=12
\relax
\hybrid
\numberbysection

\begin{document}
\title{Three-body local correlation function in the Lieb-Liniger model: bosonization approach}

\author{Vadim V. Cheianov\thanks{Department of Physics, Lancaster University, Lancaster,
LA1 4YB, United Kingdom}, H. Smith\thanks{Niels Bohr Institute,
Universitetsparken 5, DK-2100 Copenhagen, Denmark}, and M. B.
Zvonarev\thanks{DPMC, University of Geneva, Quai Ernest Ansermet 24,
1211 Geneva, Switzerland}}

\maketitle

\abstract{We develop a method for the calculation of vacuum
expectation values of local operators in the Lieb-Liniger model.
This method is based on a set of new identities obtained using
integrability and effective theory (``bosonization'') description.
We use this method to get an explicit expression for the three-body
local correlation function, measured in a recent experiment
\cite{Tolra}.}

\section{Introduction}

In this paper we present an approach to the calculation of the
vacuum expectation values of local operators in the Lieb-Liniger
model. Although the technique is general we shall concentrate on an
particular operator, $\psi^\dagger(x)^3\psi(x)^3,$ whose vacuum
expectation value was measured in a recent experiment~\cite{Tolra}.
The physical motivation for the study of this problem and the
analysis of the resulting expression has been given in
Ref.~\cite{CSZ-05}. Here we focus on the mathematical aspects of the
problem and present the details of the derivation. In some
integrable models used in condensed matter physics the exact
expressions for expectation values of local operators are known,
Refs.~\cite{S-92,LZ-97}. However, it is not known how to use these
results for the calculations in the Lieb-Liniger model.

The Lieb-Liniger model describes a one-dimensional gas of bosons
interacting via a $\delta$-potential~\cite{LL-63}. Its Hamiltonian
is
\begin{equation}
H=\int_0^{L} dx \left[-\psi^{\dagger}(x) \partial_x^2
\psi(x)+ c \psi^\dagger(x)\psi^\dagger(x) \psi(x)\psi(x)\right],
\label{qContHam}
\end{equation}
where the units have been chosen such that $\hbar=2m=1.$ The boson
fields $\psi$ and $\psi^\dagger$ satisfy canonical equal-time
commutation relations:
\begin{equation}
[\psi(x),\psi^\dagger(y)]=\delta(x-y), \qquad
[\psi^\dagger(x),\psi^\dagger(y)]=[\psi(x),\psi(y)]=0.
\label{commpsi}
\end{equation}
The system is placed on a ring of circumference $L$ and periodic
boundary conditions are imposed:
\begin{equation}
\psi^\dagger(0)=\psi^\dagger(L), \qquad \psi(0)=\psi(L).
\end{equation}
The particle number operator $N$ is
\begin{equation}
N=\int_0^L dx\,\psi^\dagger(x)\psi(x), \label{Ndef}
\end{equation}
and the momentum operator $P$ is
\begin{equation}
P=\frac{i}2\int_0^L
dx\left\{\left[\partial_x\psi^\dagger(x)\right]\psi(x)-
\psi^\dagger(x)\partial_x\psi(x)\right\}. \label{Pdef}
\end{equation}
These operators are integrals of motion:
\begin{equation}
[H,N]=[H,P]=0.
\end{equation}
The interaction constant $c\ge0$ has the dimension of inverse
length. The dimensionless coupling strength $\gamma$ is given by
\begin{equation}
\gamma=\frac{c}D, \label{gamma}
\end{equation}
where $D$ is the particle density,
\begin{equation}
D=\frac{N}L. \label{ddef}
\end{equation}

The model \eqref{qContHam} has been studied extensively. Its exact
eigenfunctions, spectrum, and thermodynamics were obtained during
1960s, while the calculation of its correlation functions still
remains a challenge (for a review of the field see, for example,
Ref.~\cite{KBI-93}). We obtain in the present paper an exact
expression for the local correlation function
\begin{equation}
g_3(\gamma)=\langle\psi^\dagger(x)^3\psi(x)^3\rangle, \label{g3def}
\end{equation}
where $\langle\cdots\rangle$ denotes the expectation value in the
ground state $|\mathrm{gs}\rangle$ of the system\footnote{Since the
ground state is translationally invariant, $g_3$ do not depend on
$x$. The matrix elements, in particular, the expectation values, of
local operators, like $\psi^\dagger(x)^3\psi(x)^3,$ are often called
form-factors.},
\begin{equation}
\langle\cdots\rangle\equiv
\langle\mathrm{gs}|\cdots|\mathrm{gs}\rangle. \label{gsmic}
\end{equation}
Let us formulate the main result: in the limit of the infinite
number of particles taken at a finite density,
\begin{equation}
D\to\mathrm{const}\ne0,\infty \qquad \text{as} \qquad N,L\to\infty,
\label{tl}
\end{equation}
the correlation function \eqref{g3def} is expressed as
\begin{equation}
\frac{g_3(\gamma)}{n^3}=
\frac{3}{2\gamma}\epsilon_4'-\frac{5\epsilon_4}{\gamma^2}
+\left(1+\frac{\gamma}{2}\right)\epsilon_2'-2\frac{\epsilon_2}{\gamma}-
\frac{3\epsilon_2\epsilon_2'}{\gamma}+\frac{9\epsilon_2^2}{\gamma^2},
\label{g3result}
\end{equation}
where $\epsilon_m$ are moments of the quasi-momentum distribution
function $\sigma(k)$
\begin{equation}
\epsilon_m=\left(\frac{\gamma}{\alpha}\right)^{m+1}
\int_{-1}^{1}dz\,z^m\sigma(z), \label{b3}
\end{equation}
and $\epsilon_m^\prime$ is the derivative of $\epsilon_m$ with
respect to $\gamma.$ The function $\sigma(k)$ is the solution to the
linear integral equation
\begin{equation}
\sigma(z)-\frac1{2\pi}\int_{-1}^1dy\,\frac{2\alpha\sigma(y)}
{\alpha^2+(z-y)^2}=\frac1{2\pi}, \label{b1}
\end{equation}
where $\alpha$ is an implicit function of $\gamma:$
\begin{equation}
\alpha=\gamma\int_{-1}^1dz\,\sigma(z) \label{b2}.
\end{equation}

The paper is organized as follows: In section \ref{sect:qblm} we
consider the $q$-boson lattice model. This is an integrable lattice
model whose continuum limit is the Lieb-Liniger model. The studies
of the Noether currents in the $q$-boson lattice and the
Lieb-Liniger model are performed in section \ref{sect:noether}. In
section \ref{sect:bosonization} the $q$-boson lattice model is
bosonized. In section \ref{sect:stringid} we write some
contour-independent integral identities relating short- and
long-distance properties of correlation functions in the $q$-boson
lattice model. At long distances the results of
section~\ref{sect:bosonization} are applicable. We thus get some
non-trivial information about the short-distance properties of
correlation functions. In section \ref{sect:3bcf} we take the
continuum limit of expressions obtained in section
\ref{sect:stringid}. This leads us to Eq.~\eqref{g3result} for
$g_3(\gamma).$ This equation and the contour-independent integral
identities of section \ref{sect:stringid} are the main results of
our paper.

\section{$q$-boson lattice model \label{sect:qblm}}

In this section a lattice regularization of the Lieb-Liniger model
\eqref{qContHam} is described. Lattice regularization is a natural
tool to circumvent the short-range singularities discussed in
section \ref{sect:NCLL} and can, in principle, be done in many
different ways. An integrable lattice regularization will be needed
for our purposes. Several different regularization schemes
preserving integrability are discussed in Ref.~\cite{AK-04}. We
shall use the so-called $q$-boson lattice model \cite{BBP-93,BIK-98}
as an integrable lattice regularization of the Lieb-Liniger model.
This choice is motivated by its simplicity.

In section \ref{sect:qba} the $q$-boson algebra is constructed for a
given lattice site. In section \ref{sect:hqblm} we present the
Hamiltonian of the $q$-boson lattice model defined on a lattice with
an arbitrary number of sites, $M.$ In the continuum limit, defined
by Eq.~\eqref{qCont}, this Hamiltonian becomes the Hamiltonian of
the Lieb-Liniger model, Eq.~\eqref{qContHam}. In section
\ref{sect:imqbm} we discuss the integrability of the $q$-boson
lattice model. The generating functional of the integrals of motion
is constructed using the formalism of the Quantum Inverse Scattering
Method. By expanding this functional the explicit expressions for
several integrals of motion are obtained. In section \ref{sect:tqbm}
we study the eigenfunctions and the eigenvalues of the integrals of
motion in the $q$-boson lattice model. Finally, we apply the
Hellmann-Feynman formula to one of the integrals of motion in
section \ref{sect:HFi} with the aim of using the resulting identity
in subsequent sections.

\subsection{$q$-boson algebra \label{sect:qba}}
Consider the operators $B_n,$ $B_n^\dagger$ and $N_n=N^\dagger_n$
satisfying the so-called $q$-boson algebra
\begin{eqnarray}
& B_n B_n^{\dagger}-q^{-2} B_n^{\dagger} B_n=1,  \label{qalg0}\\
& [N_n, B_n]=-B_n, \qquad [N_n, B_n^\dagger]=B_n^\dagger.
\label{qalg}
\end{eqnarray}
The index $n$ labels the sites of a lattice; it will be held fixed
in this section, where we work with the operators for a given
lattice site. The parameter $q$ is a $c$-number and it is enough for
our purposes to work with real $q\ge1.$

Define now a Fock space for the $q$-boson algebra given by
Eqs.~\eqref{qalg0} and \eqref{qalg}. Let the basis states
$|m_j\rangle_j,$ $m_j=0,1,2,\ldots$ of the Fock space be the states
of the harmonic oscillator
\begin{equation}
b_n|m_n\rangle_n= m_n^{1/2}|m_n-1\rangle_n, \qquad
b^\dagger_n|m_n\rangle_n=(m_n+1)^{1/2}|m_n+1\rangle_n, \label{ho}
\end{equation}
where $b^\dagger_n$ and $b_n$ are the canonically commuting
creation and annihilation operators
\begin{equation}
[b_n, b^\dagger_n]= 1.
\label{latticebosons}
\end{equation}
In this Fock space the operator $N_n$ entering Eqs.~\eqref{qalg}
acts in a manner identical with that of the operator
\begin{equation}
N_n=b^\dagger_n b_n, \label{Nbdef}
\end{equation}
for which we have
\begin{equation}
N_n|0\rangle_n=0, \qquad N_n|m_n\rangle_n = m_n|m_n\rangle_n, \qquad
m_n=1,2,3,\ldots. \label{Nbeig}
\end{equation}
The operators $B^\dagger_n$ and $B_n$ entering Eqs.~\eqref{qalg0}
and \eqref{qalg} act in the Fock space as follows
\begin{equation}
B_n|m_n\rangle_n=[m_n]_q^{1/2}|m_n-1\rangle_n, \qquad
B^\dagger_n|m_n\rangle_n=[m_n+1]_q^{1/2}|m_n+1\rangle_n,
\label{Fock}
\end{equation}
where
\begin{equation}
[x]_q\equiv\frac{1-q^{-2x}}{1-q^{-2}}. \label{[x]}
\end{equation}
It can be easily shown that the commutation relations \eqref{qalg0}
and \eqref{qalg} are consistent with Eqs.~\eqref{Nbeig} and
\eqref{Fock}. Note that
\begin{equation}
[x]_q\to x \qquad \textrm{as} \qquad q\to1,
\end{equation}
and therefore
\begin{equation}
B^\dagger_n\to b^\dagger_n \qquad \textrm{and} \qquad B_n\to b_n
\qquad \textrm{as} \qquad q\to1.
\end{equation}

It should be emphasized that the present paper deals with a {\it
representation}, Eqs.~\eqref{Nbeig} and \eqref{Fock}, of the
$q$-boson algebra, but not with the algebra itself.  All statements
concerning the $q$-boson algebra should be understood as statements
about this particular representation. One can see from
Eq.~\eqref{Nbdef} that this choice of representation makes it
possible to relate the operator $N_n$ entering Eqs.~\eqref{qalg}
with the canonical boson operators $b^\dagger_n$ and $b_n.$ It is
also possible to express the operators $B^\dagger_n$ and $B_n$ in
terms of $b^\dagger_n$ and $b_n:$
\begin{equation}
B_n= \sqrt{\frac{[N_n+1]_q}{N_n+1}} b_n, \qquad B_n^{\dagger} =
b^\dagger_n\sqrt{\frac{[N_n+1]_q}{N_n+1}} \label{Bn=bn}
\end{equation}
and give an alternative form of the commutation relation
\eqref{qalg0}:
\begin{equation}
[B_n,B^\dagger_n]=q^{-2N_n}.\label{BBalt}
\end{equation}
The relations \eqref{Bn=bn} and \eqref{BBalt} can be easily proven
using Eqs.~\eqref{ho} and \eqref{Fock}.

We see from Eqs.~\eqref{Nbdef} and \eqref{Bn=bn} that the relations
between the $q$-boson operators $B^\dagger_n$ and $B_n$ and
canonical boson operators $b^\dagger_n$ and $b_n$ are nontrivial.
Furthermore, it follows from Eqs.~\eqref{Fock} that
\begin{equation}
B_n^\dagger B_n |m_n\rangle_n = [m_n]_q|m_n\rangle_n,
\end{equation}
and therefore that
\begin{equation}
B_n^\dagger B_n=[N_n]_q=\frac{1-q^{-2N_n}}{1-q^{-2}}. \label{BBviaN}
\end{equation}
Thus, $B^\dagger_nB_n\ne N_n,$  and $N_n$ is non-polynomial in terms
of $B^\dagger_n$ and $B_n.$

\subsection{Hamiltonian of the $q$-boson lattice model \label{sect:hqblm}}

In order to pass from the quantum mechanical (one-site) model
considered in the previous section to a quantum field theory model
we introduce a lattice with $M$ sites; the index $n$ in $B_n,$
$B^\dagger_n$ and $N_n$ labels the lattice sites. We impose an
``ultralocality'' condition on the operators $B_{n_1},$
$B_{n_2}^\dagger$ and $N_{n_3}$ by requiring that they commute at
different lattice sites. The basis states of the whole lattice are
constructed as the tensor product of the local basis states:
\begin{equation}
|0\rangle=\otimes_{j=1}^M|0\rangle_j, \qquad
|m\rangle=\otimes_{j=1}^M|m_j\rangle_j.\label{MFock}
\end{equation}

The Hamiltonian for the $q$-boson lattice model is defined as
follows:
\begin{equation}
H_q=-\frac{1}{\delta^2} \sum_{n=1}^{M}
(B_n^{\dagger}B_{n+1}+B_{n+1}^\dagger B_n  - 2 N_n) \label{qHam}
\end{equation}
with the periodic boundary conditions $M+1=1$ imposed. Here $\delta$
is the lattice spacing. The factor $\delta^{-2}$ is introduced to
ensure the proper continuum limit. One can check using the $q$-boson
algebra, Eqs.~\eqref{qalg0} and \eqref{qalg}, that $H_q$ commutes
with the number operator
\begin{equation}
N=\sum_{n=1}^M N_n.\label{N}
\end{equation}
Since $N$ is non-polynomial in terms of $B^\dagger_n$ and $B_n,$
which can be seen from Eq.~\eqref{BBviaN}, $H_q$ is non-polynomial
in terms of these fields. It is non-polynomial in terms of the
canonical lattice bosons $b^\dagger_n$ and $b_n$ as well, now due to
the non-polynomiality of the hopping term. Thus the model is
interacting and the interaction is encoded in the deformation
parameter $q:$ if one takes $q\to 1$ the result will be a quadratic
boson Hamiltonian on a lattice, describing nearest-neighbor hopping:
\begin{equation}
H_q\to -\frac1{\delta^2}\sum_{n=1}^M
(b^\dagger_nb_{n+1}+b^\dagger_{n+1}b_n-2N_n) \qquad \text{as} \qquad
q\to1.
\end{equation}

The limit $q\to1$ is somehow trivial. It is much more fruitful to
consider the following limit: let $\delta\to0,$ $M\to\infty,$ and
$q\to1,$ while $L$ and $c$ are kept constant:
\begin{equation}
L=M\delta, \qquad c/2=\kappa\delta^{-1}, \qquad \text{as}
\qquad\delta\to 0, \qquad M\to\infty, \qquad \kappa\to0,
\label{qCont}
\end{equation}
where $\kappa$ is related to $q$ as follows
\begin{equation}
q=e^{\kappa}. \label{qgamma}
\end{equation}
The limit \eqref{qCont} will be called the {\it continuum limit} of
the $q$-boson lattice model. The sums in the continuum limit are
converted into integrals in the usual way
\begin{equation}
\delta\sum_{n=1}^M\to\int_0^L dx.
\end{equation}
For any $q$-deformed quantity $[x]_q$, Eq.~\eqref{[x]}, the
following expansion is valid
\begin{equation}
[x]_q=x-\kappa x(x-1)+\frac{\kappa^2}{3}x(x-1)(2x-1)+\dots, \qquad
\kappa\to0, \label{xqexp}
\end{equation}
where $q$ is related to $\kappa$ by Eq.~\eqref{qgamma}. Therefore
\begin{equation}
\sqrt{\frac{[N_n+1]_q}{N_n+1}}=
1-\frac{\kappa}2N_n+\frac{\kappa^2}{24}N_n(5N_n+4)+\dots, \qquad
\kappa\to0.
\end{equation}
We define the continuum boson fields $\psi(x)$ and $\psi^\dagger(x)$
by
\begin{equation}
\psi^\dagger(x)= \delta^{-1/2}b^\dagger_{n}, \qquad \psi(x)=
\delta^{-1/2}b_{n},
\end{equation}
where
\begin{equation}
x=\delta n. \label{xdn}
\end{equation}
The fields $\psi^\dagger(x)$ and $\psi(x)$ satisfy canonical
commutation relations \eqref{commpsi}. One can easily check that the
$q$-boson Hamiltonian \eqref{qHam} becomes the Hamiltonian of the
Lieb-Liniger model \eqref{qContHam} in the continuum limit
\eqref{qCont}.

Finally, we discuss the number of particles and momentum operators
in the $q$-boson lattice model and their continuum limit. We have
defined the number operator in the $q$-boson lattice model by
Eq.~\eqref{N}. The continuum limit \eqref{qCont} of this operator is
given by Eq.~\eqref{Ndef}. The momentum operator is defined in the
$q$-boson lattice model as follows:
\begin{equation}
P_q=-\frac{i}2\frac1{\delta}\sum_{n=1}^M(B^\dagger_n
B_{n+1}-B^\dagger_{n+1}B_n). \label{Pqdef}
\end{equation}
Its continuum limit is given by Eq.~\eqref{Pdef}. To avoid any
confusion, we note that $P_q$ is not a generator of lattice
translations, that is
\begin{equation}
e^{-iP_q}B_ne^{iP_q}\ne B_{n+1}.
\end{equation}
However, its continuum limit \eqref{Pdef} is such a generator for
the continuum model:
\begin{equation}
e^{-iaP}\psi(x)e^{iaP}= \psi(x+a). \label{tgen}
\end{equation}
It is due to the property \eqref{tgen} that we call $P_q$ the
momentum operator. We shall not need to know an explicit form of the
true momentum operator $P^{\mathrm{true}}_q$ which generates the
lattice translations in the $q$-boson lattice model.

\subsection{Integrals of motion in the $q$-boson lattice model \label{sect:imqbm}}
An infinite set of integrals of motion in the $q$-boson lattice
model can be constructed using the Quantum Inverse Scattering
Method. A detailed description of this method is given, for example,
in Ref.~\cite{KBI-93}, and it was applied to the $q$-boson lattice
model, in particular, in Refs.~\cite{BBP-93,BIK-98}. We give in the
present section a schematic description of the method, referring the
reader to the Refs.~\cite{BBP-93,BIK-98,KBI-93} for details. The
main purpose of the section is to obtain
Eqs.~\eqref{Imtilde}--\eqref{J3}.

The so-called $L$-operator for the $q$-boson lattice model is
defined by
\begin{equation}
L_{n}(\lambda)= \left(
\begin{array}{cc}
e^{\lambda} & \chi B_n^\dagger \\
\chi B_n & e^{-\lambda}
\end{array}
\right), \label{qL}
\end{equation}
where
\begin{equation}
\chi=\sqrt{1-q^{-2}}=\sqrt{1-e^{-2\kappa}}. \label{chi}
\end{equation}
The $L$-operator \eqref{qL} is a $2\times2$ matrix with the entries
being quantum operators acting in the infinite-dimensional Fock
space defined by Eq.~\eqref{ho}. The Quantum Inverse Scattering
Method is based on the existence of the intertwining relation for
the $L$-operator:
\begin{equation}
R(\lambda, \mu) \left(L_n(\lambda)\otimes L_n(\mu) \right)=
\left(L_n(\mu)\otimes L_n(\lambda) \right) R(\lambda, \mu)
\label{qBilinear}
\end{equation}
with the $R$-matrix defined by
\begin{equation}
R(\lambda, \mu)= \left(
\begin{array}{cccc}
f(\mu, \lambda) &0&0&0 \\
0&g(\mu, \lambda) &q&0\\
0&q^{-1}&g(\mu, \lambda)&0\\
0&0&0&f(\mu, \lambda)
\end{array}
\right), \label{qR}
\end{equation}
where
\begin{equation}
f(\lambda,
\mu)=\frac{\sinh(\lambda-\mu+\kappa)}{\sinh(\lambda-\mu)}, \qquad
g(\lambda, \mu)=\frac{\sinh(\kappa)}{\sinh(\lambda-\mu)}.
\label{fgdef}
\end{equation}
Recall that the tensor product for $2\times2$ matrices $a$ and $b$
is defined as follows:
\begin{equation}
a\otimes b=\left(
\begin{array}{cc}
a_{11}&a_{12}\\
a_{21}&a_{22}
\end{array}
\right)\otimes \left(
\begin{array}{cc}
b_{11}&b_{12}\\
b_{21}&b_{22}
\end{array}
\right)=\left(
\begin{array}{cccc}
a_{11}b_{11} &a_{11}b_{12}&a_{12}b_{11}&a_{12}b_{12}\\
a_{11}b_{21}&a_{11}b_{22}&a_{12}b_{21}&a_{12}b_{22}\\
a_{21}b_{11}&a_{21}b_{12}&a_{22}b_{11}&a_{22}b_{12}\\
a_{21}b_{21}&a_{21}b_{22}&a_{22}b_{21}&a_{22}b_{22}
\end{array}
\right).
\end{equation}
The monodromy matrix $T(\lambda)$ is defined as a matrix product of
the $L$-operators taken over all lattice sites
\begin{equation}
T(\lambda)=\left(
\begin{array}{cc}
A(\lambda) & B(\lambda) \\
C(\lambda) & D(\lambda)
\end{array}
\right)=L_M(\lambda)L_{M-1}(\lambda)\cdots L_1(\lambda).
\label{Tdef}
\end{equation}
The entries $A(\lambda),\ldots,D(\lambda)$ of the monodromy matrix
are quantum operators acting in the tensor product of the local Fock
spaces over all sites of the lattice. Due to the relation
\eqref{qBilinear} and the commutativity of the entries of the
$L$-operator \eqref{qL} at different lattice sites one has the
intertwining relation for the monodromy matrix
\begin{equation}
R(\lambda, \mu) \left(T(\lambda)\otimes T(\mu) \right)=
\left(T(\mu)\otimes T(\lambda) \right) R(\lambda, \mu).
\label{qTBilinear}
\end{equation}
Equation \eqref{qTBilinear} defines $16$ commutation relations for
the operators entering the monodromy matrix. We write explicitly
those relations which we shall use in deriving the Bethe equations
in section \ref{sect:tqbm}:
\begin{align}
&qA(\lambda)B(\mu)=f(\lambda,\mu)B(\mu)A(\lambda)+g(\mu,\lambda)B(\lambda)A(\mu),\\
&qD(\lambda)B(\mu)=f(\mu,\lambda)B(\mu)D(\lambda)+g(\lambda,\mu)B(\lambda)D(\mu),\\
&[B(\lambda),B(\mu)]=0.
\end{align}

The transfer matrix $\tau(\lambda)$ is defined as the trace over the
matrix space of the monodromy matrix
\begin{equation}
\tau(\lambda)=\mathrm{Tr}\, T(\lambda)=A(\lambda)+D(\lambda).
\label{taudef}
\end{equation}
It can be proven (see chapter VI of Ref.~\cite{KBI-93} for details)
that for any $\lambda$ and $\mu$
\begin{equation}
[\tau(\lambda), \tau(\mu)]=0 \label{taucom}
\end{equation}
which implies that $\tau$ is a generating function of the
integrals-of-motion of the problem: expanding $\tau(\lambda)$ in
$\lambda$ one gets a set of commuting integrals-of-motion $I_m$.
This set can be chosen in many different ways since any analytic
function of $\tau(\lambda)$ can play the role of the generating
functional. We, however, impose an additional very restrictive
locality condition on these integrals by requiring them to be
written in the following form
\begin{equation}
I_m=\delta\sum_{n=1}^M \mathcal{J}^{(m)}_\tau(n), \label{IJ}
\end{equation}
where the operators $\mathcal{J}^{(m)}_\tau(n)$ act nontrivially in
$m$ neighboring lattice sites only. The subscript $\tau$ appears in
$\mathcal{J}^{(m)}_\tau(n)$ since these local operators are
$\tau$-components of the corresponding Noether currents, which will
be discussed in section \ref{sect:noether}. To get the set $\{I_m\}$
we introduce the variable
\begin{equation}
\zeta=e^{\lambda} \label{xi->lambda}
\end{equation}
and consider the expression
\begin{equation}
I_{m}= \left.\frac{1}{(2m)!} \frac{d^{2m}}{d \zeta^{2m}}
\ln\left[\zeta^M\tau(\zeta)\right] \right|_{\zeta\to 0}, \qquad
m=1,2,3,\ldots.\label{Imtilde}
\end{equation}
We assume that $\lambda$ is real, therefore $\zeta$ is real and
nonnegative. The local operators $\mathcal{J}_\tau^{(1)}(n),$
$\mathcal{J}_\tau^{(2)}(n)$ and $\mathcal{J}_\tau^{(3)}(n)$ are
\begin{align}
\label{J1}
&\mathcal{J}_\tau^{(1)}(n)=\frac{1}{\delta}\chi^2 B^\dagger_n B_{n+1} \\
\label{J2} &\mathcal{J}_\tau^{(2)}(n)=
\frac{1}{\delta}\chi^2\left(1-\frac{\chi^2}{2}\right) \left(
B^\dagger_n B_{n+2} - \frac{\chi^2}{2-\chi^2} B_n^\dagger
B_n^\dagger B_{n+1}B_{n+1}- \chi^2 B_n^\dagger B_{n+1}^\dagger
B_{n+1}B_{n+2} \right)
\end{align}
and
\begin{multline}
\mathcal{J}_\tau^{(3)}(n)= \frac1\delta\chi^2
\left(1-\chi^2+\frac{\chi^4}{3} \right)
\left(\vphantom{\frac{\chi^4}{3}} B^\dagger_n B_{n+3} -\chi^2
B^\dagger_nB^\dagger_n B_{n+1}B_{n+2}-
 \chi^2 B^\dagger_n B^\dagger_{n+1} B_{n+1}B_{n+3} \right.\\
- \chi^2 B^\dagger_n B^\dagger_{n+1} B_{n+2}B_{n+2}- \chi^2
B^\dagger_n B^\dagger_{n+2} B_{n+2}B_{n+3} + \frac{\chi^4}{3-3
\chi^2+\chi^4}B^\dagger_n B^\dagger_n B^\dagger_n
B_{n+1}B_{n+1}B_{n+1}
\\
+\chi^4 B^\dagger_n B^\dagger_n B^\dagger_{n+1}
B_{n+1}B_{n+1}B_{n+2}+ \chi^4 B^\dagger_n B^\dagger_{n+1}
B^\dagger_{n+1} B_{n+1}B_{n+2}B_{n+2}
\\
+ \left. \chi^4 B^\dagger_n B^\dagger_{n+1} B^\dagger_{n+2}
B_{n+1}B_{n+2}B_{n+3} \vphantom{\frac{\chi^4}{3}} \right) \label{J3}
\end{multline}
respectively. To calculate $g_3(\gamma),$ Eq.~\eqref{g3def}, we will
need $\mathcal{J}_\tau^{(1)}$ and $\mathcal{J}_\tau^{(2)}$; the
explicit expression for $\mathcal{J}_\tau^{(3)}$ is displayed in
order to illustrate how the complexity of $\mathcal{J}_\tau^{(m)}$
grows with increasing $m$.

The integrals $I_m$ are non-Hermitian, they contain both real and
imaginary part. Using the involution
\begin{equation}
[\tau(\zeta)]^\dagger=\tau(\zeta^{-1}) \label{invo}
\end{equation}
it may be shown that
\begin{equation}
[I^{\dagger}_m, I_n]=0 \qquad \text{for any} \quad m, n.
\label{IdIcomm}
\end{equation}
For the integrals of motion $I^\dagger_m,$ we use the following
notation
\begin{equation}
I_{-m}\equiv I^\dagger_m, \qquad m=1,2,3,\ldots, \label{I-m}
\end{equation}
and for the local operators $\mathcal{J}_\tau$
\begin{equation}
\mathcal{J}^{(-m)}_\tau(n)\equiv[\mathcal{J}^{(m)}_\tau(n)]^\dagger,
\qquad m=1,2,3,\dots. \label{J-m}
\end{equation}
Using the involution \eqref{invo} one gets for $I_{-m}$
\begin{equation}
I_{-m}= \left.\frac{1}{(2m)!} \frac{d^{2m}}{d \zeta^{2m}}
\ln\left[\zeta^M\tau(\zeta^{-1})\right] \right|_{\zeta\to 0}, \qquad
m=1,2,3,\ldots.\label{Imd}
\end{equation}
A set of common eigenfunctions of $I_m,$ $m=\pm1,\pm2,\dots$ is
constructed in section \ref{sect:tqbm}.

The local operators $\mathcal{J}_\tau^{(m)}(n)$ generated by
Eqs.~\eqref{IJ} and \eqref{Imtilde} are polynomial in $B^\dagger_n$
and $B_n,$ while the one-site number operator $N_n$ is
non-polynomial in these variables, as was noticed below
Eq.~\eqref{BBviaN}. Therefore, the number operator $N,$
Eq.~\eqref{N}, cannot be expressed as a finite linear combination of
the integrals of motion $I_m,$ Eq.~\eqref{Imtilde}. It is, however,
clear from the structure of $\mathcal{J}_\tau^{(m)}(n)$ that
\begin{equation}
[N,\mathcal{J}_\tau^{(m)}(n)]=0, \qquad m=\pm1,\pm2,\dots .
\label{NJtaucomm}
\end{equation}
Indeed, $N$ commutes with any monomial containing an equal number of
the creation and annihilation operators, $B^\dagger$ and $B.$ It
follows from Eq.~\eqref{NJtaucomm} that
\begin{equation}
[N,I_m]=0, \qquad m=\pm1,\pm2,\dots.
\end{equation}
To stress that $N$ is one of the integrals of motion, we shall use
the following notation
\begin{equation}
N\equiv I_0, \qquad \mathcal{J}_\tau^{(0)}(n)\equiv \frac1\delta
N_n. \label{NJ0}
\end{equation}
The Hamiltonian \eqref{qHam} can thus be written as
\begin{equation}
H_q=-\frac1{\chi^2\delta^2}(I_1+I_{-1}-2\chi^2 I_0), \label{HviaI}
\end{equation}
and the momentum operator \eqref{Pqdef} as
\begin{equation}
P_q=-\frac{i}2 \frac{1}{\chi^2\delta}(I_1-I_{-1}).
\end{equation}

\subsection{Eigenfunctions and eigenvalues of the integrals of motion
in the $q$-boson lattice model \label{sect:tqbm}}
We have constructed in section \ref{sect:imqbm} a set of integrals
of motion $I_m,$ Eq.~\eqref{Imtilde}, of the $q$-boson lattice model
on a lattice with an arbitrary number of sites, $M$. We find in the
present section their common eigenfunctions and their eigenvalues
using the Algebraic Bethe Ansatz technique \cite{KBI-93}, an
important ingredient of the Quantum Inverse Scattering Method. The
eigenvalues of all $I_m$ are defined in the $N$-particle sector by
$N$ parameters called quasi-momenta. These quasi-momenta are the
solutions of the system of nonlinear equations called the Bethe
equations. Following the presentation of Ref.~\cite{KBI-93},
Chap.~I, we find from the analysis of the Bethe equations the ground
state quasi-momentum distribution and study its properties in the
limit defined by Eq.~\eqref{lattlim}.

The cornerstone of the Algebraic Bethe Ansatz method is the fact
that the vacuum $|0\rangle$ defined by Eq.~\eqref{MFock} annihilates
the $C(\lambda)$ entry of the monodromy matrix \eqref{Tdef} and is
an eigenfunction for the $A(\lambda)$ and $D(\lambda)$ entries:
\begin{equation}
A(\lambda)|0\rangle=e^{M\lambda}|0\rangle, \qquad
D(\lambda)|0\rangle=e^{-M\lambda}|0\rangle.
\end{equation}
Since $N$ is a good quantum number, one can work in the $N$-particle
sector. Define a set of states by the following formula
\begin{equation}
|\psi_N(\lambda_1,\ldots,\lambda_N)\rangle=
\prod_{j=1}^{N}B(\lambda_j)|0\rangle.\label{Bethestates}
\end{equation}
These states (often called the Bethe states) are eigenfunctions of
the transfer matrix \eqref{taudef} and, hence, of the integrals of
motion $I_m,$ $m=\pm1,\pm2,\dots,$ if the parameters
$\lambda_1,\ldots,\lambda_M$ satisfy a system of coupled nonlinear
equations (called the Bethe equations),
\begin{equation}
e^{2M\lambda_j}=\prod_{\genfrac{}{}{0pt}{1}{k=1}{k\ne j}}^N
\frac{f(\lambda_k,\lambda_j)}{f(\lambda_j,\lambda_k)}, \qquad
j=1,\ldots,N.\label{Bel}
\end{equation}
The eigenvalues $\theta_N$ of the transfer matrix acting on the
Bethe states \eqref{Bethestates} are given by
\begin{align}
&\tau(\lambda)|\psi_N\rangle=
\theta_N(\lambda,\{\lambda_j\})|\psi_N\rangle, \label{tsd}\\
&q^N\theta_N(\lambda,\{\lambda_j\})=e^{M\lambda}\prod_{j=1}^N
f(\lambda,\lambda_j)+e^{-M\lambda}\prod_{j=1}^N
f(\lambda_j,\lambda).\label{tauspectrum}
\end{align}
The eigenvalues of the integrals of motion $I_{m},$
$m=\pm1,\pm2,\dots$ acting on the Bethe states \eqref{Bethestates}
can be obtained by acting with the representations \eqref{Imtilde}
and \eqref{Imd} onto $|\psi_N\rangle$ and using
Eqs.~\eqref{xi->lambda}, \eqref{tsd} and \eqref{tauspectrum}. The
calculations are tedious, while the final result is surprisingly
simple:
\begin{equation}
I_m|\psi_N\rangle=\left(1-q^{-2|m|}\right)\frac1{|m|}\sum_{j=1}^N
e^{-2m\lambda_j}|\psi_N\rangle, \qquad m=\pm1,\pm2,\dots.
\label{Imeig}
\end{equation}
It is useful to mention that, in order to calculate the correlation
function \eqref{g3def} we shall need to know the spectrum of the
integrals $I_{\pm1}$ and $I_{\pm2}$ only. For $|m|>2$ we did not
obtain Eq.~\eqref{Imeig} analytically. Instead, we simply checked
that it is correct for some given values of $N,$ $M$ and $m$ using
the \textsf{Mathematica} package.

It will be convenient to use instead of $\lambda_1,\ldots,\lambda_N$
a set of quasi-momenta $p_1,\ldots,p_N:$
\begin{equation}
\lambda_j=i\frac{p_j}2, \qquad j=1,\ldots,N. \label{lp}
\end{equation}
Written in these variables, the Bethe equations \eqref{Bel} are
\begin{equation}
e^{iMp_j}=\prod_{\genfrac{}{}{0pt}{1}{k=1}{k\ne j}}^N
\frac{\sin[\frac12(p_j-p_k)+i\kappa]}{\sin[\frac12(p_j-p_k)-i\kappa]},
\qquad j=1,\ldots,N. \label{Be2}
\end{equation}
Using Eqs.~\eqref{HviaI}, \eqref{Imeig} and \eqref{lp}, one gets for
the eigenvalues of the Hamiltonian \eqref{qHam}:
\begin{equation}
E_N=\frac4{\delta^2}\sum_{j=1}^N\sin^2\frac{p_j}2.\label{EN}
\end{equation}

We now discuss some properties of the Bethe equations necessary to
identify the ground state of the model and to take the limit
\eqref{lattlim}. The analysis will be very similar to that one
carried out for the Lieb-Liniger model in Ref.~\cite{KBI-93},
Chap.~I, so we omit several long proofs, referring the reader to
Ref.~\cite{KBI-93} for details.

(i) All the solutions $p_j$ of the Bethe equations \eqref{Be2} are
real. The proof is the same as that given for the Lieb-Liniger model
in Ref.~\cite{KBI-93}, page 11.

(ii) It follows from Eq.~\eqref{EN} that $E_N$ is a periodic
function of the quasi-momenta $p_j$ with period $2\pi.$ We shall
work with quasi-momenta lying in the interval
\begin{equation}
 -\pi<p_j<\pi, \qquad j=1,\ldots,N. \label{prange}
\end{equation}
The condition \eqref{prange} will be assumed in all subsequent
formulas.

(iii) We write the Bethe equations \eqref{Be2} in the logarithmic
form
\begin{equation}
Mp_j+\sum_{k=1}^N \theta(p_j-p_k)=2\pi\left(n_j+\frac{N-1}2\right),
\qquad j=1,\ldots,N, \label{Be3}
\end{equation}
where the parameters $n_j$ take arbitrary integer values:
\begin{equation}
n_j \text{ is an arbitrary integer}, \qquad j=1,\dots,N. \label{n_j}
\end{equation}
From now on we shall work with odd $N:$
\begin{equation}
N \text{ is odd}.
\end{equation}
The function $\theta(p)$ is
\begin{equation}
\theta(p)=i\ln\frac{\sin\left(i\kappa+\frac12p\right)}
{\sin\left(i\kappa-\frac12p\right)}, \qquad \theta(-p)=-\theta(p).
\end{equation}
The derivative of $\theta(p)$ is positive,
\begin{equation}
\theta^\prime(p)=\frac{i\sin(2i\kappa)}
{2\sin(i\kappa+\frac12p)\sin(i\kappa-\frac12p)}=
\frac{\sinh(2\kappa)}{\cosh(2\kappa)-\cos p}>0, \label{tprime}
\end{equation}
therefore $\theta(p)$ grows monotonously in the interval
$-\pi<p<\pi$ (recall that the condition \eqref{prange} is assumed).

(iv) For any set $\{n_j\},$ Eq.~\eqref{n_j}, there exists a uniquely
defined set $\{p_j\}$ of solutions of the Bethe equations
\eqref{Be3}. The proof is the same as that given for the
Lieb-Liniger model in Ref.~\cite{KBI-93}, page 12.

(v) We write, using Eq.~\eqref{Be3},
\begin{equation}
M(p_j-p_s)+\sum_{k=1}^N\left[\theta(p_j-p_k)-\theta(p_s-p_k)\right]=
2\pi(n_j-n_s), \qquad j,s=1,\ldots,N. \label{thetajs}
\end{equation}
Since $\theta^\prime(p)>0,$ Eq.~\eqref{tprime}, the left hand side
of the Eq.~\eqref{thetajs} is a monotonically growing function of
the parameter $p_j-p_k.$ Therefore, if $n_j>n_s$ then $p_j>p_s;$ if
$n_j=n_s$ then $p_j=p_s.$ We have thus shown that the set of
quasi-momenta $\{p_j\}$ is uniquely characterized by the set
$\{n_j\},$ and vice versa.

(vi) Like for the Lieb-Liniger model (Ref.~\cite{KBI-93}, page 14)
the energy functional \eqref{EN} taken on the sets $\{p_j\}$ of the
solutions of the Bethe equations has the minimum in the sector with
the fixed number of particles, $N,$ if $n_j$ take the values
\begin{equation}
n_j=-N+j, \qquad j=1,\ldots,N.
\end{equation}
The Bethe equations \eqref{Be3} for the ground state are, therefore,
\begin{equation}
Mp_j+\sum_{k=1}^N \theta(p_j-p_k)=2\pi\left(j-1-\frac{N-1}2\right),
\qquad j=1,\ldots,N. \label{Be4}
\end{equation}
The ground state is obviously non-degenerate.

(vii) It is obvious that the set $\{p_j\}$ of the solutions of
Eq.~\eqref{Be4} is symmetric with respect to zero. It follows then
from Eqs.~\eqref{Imeig} and \eqref{lp} that the eigenvalues of
$I_m,$ $m=\pm1,\pm2,\dots$ corresponding to the ground state wave
function $|\mathrm{gs}\rangle$ are real, and
\begin{equation}
\langle I_m\rangle=\langle I_{-m}\rangle, \qquad m=1,2,3,\dots
\label{Imgs}
\end{equation}
where $I_{-m}$ is defined by Eq.~\eqref{I-m}.

(viii) The ground state wave function is translationally invariant.
This implies that the ground state average
$\langle\mathcal{J}_\tau^{(m)}(n)\rangle$ is $n$-independent. Using
this property, one gets from Eq.~\eqref{IJ}
\begin{equation}
\langle I_m\rangle=\delta M \langle\mathcal{J}_\tau^{(m)}\rangle.
\label{Imtrans}
\end{equation}
Comparing this with Eq.~\eqref{Imgs}, one arrives at
\begin{equation}
\langle \mathcal{J}^{(-m)}_\tau \rangle= \langle
\mathcal{J}^{(m)}_\tau \rangle, \label{Jmgs}
\end{equation}
where $\mathcal{J}_\tau^{(-m)}$ is defined by Eq.~\eqref{J-m}. We
recall that the size of the system, $L,$ is the product of the
number of the sites, $M,$ and the lattice spacing, $\delta,$
Eq.~\eqref{qCont}, therefore the particle density, Eq.~\eqref{ddef},
can be written as follows
\begin{equation}
D=\frac{\langle I_0\rangle}{L}=\langle
\mathcal{J}_\tau^{(0)}\rangle. \label{rhodef}
\end{equation}

We are interested in the ground-state properties of the $q$-boson
lattice model in the limit
\begin{equation}
N\to\infty, \qquad M\to\infty.  \label{lattlim}
\end{equation}
We introduce the quasi-momentum distribution function $\rho(p_j)$ by
means of the following identity
\begin{equation}
\sum_{j=1}^N = M\sum_{j=1}^N \rho(p_j)(p_{j+1}-p_j),
\label{sum->int}
\end{equation}
where
\begin{equation}
\rho(p_j)=\frac1{M(p_{j+1}-p_j)}.
\end{equation}
The quasi-momenta $p_j$ fill the symmetric interval $-\Lambda\le
p_j\le \Lambda$ and one has $p_{j+1}-p_j\sim M^{-1}$
(Ref.~\cite{KBI-93}, page 14). For an arbitrary function $f(p_j)$
one has
\begin{equation}
\sum_{j=1}^N f(p_j)\to M\int_{-\Lambda}^\Lambda dp\, \rho(p)f(p).
\end{equation}
The parameter $\Lambda$ plays a role analogous to that of the Fermi
momentum: all states with $|p|<\Lambda$ are occupied, and all states
with $|p|>\Lambda$ are empty. The value of $\Lambda$ is defined by
the normalization condition
\begin{equation}
\frac{N}M=\int_{-\Lambda}^\Lambda dp\, \rho(p). \label{qnorm}
\end{equation}
The Bethe equations \eqref{Be4} become in the limit \eqref{lattlim}
a linear integral equation for the ground state quasi-momentum
distribution $\rho(p)$
\begin{equation}
\rho(p)-\frac1{2\pi}\int_{-\Lambda}^\Lambda d\tilde p\,K(p-\tilde
p)\rho(\tilde p)=\frac1{2\pi}, \label{Lieb}
\end{equation}
with the kernel $K(p)$ given by Eq.~\eqref{tprime},
\begin{equation}
K(p)=\theta^\prime(p)= \frac{\sinh(2\kappa)}{\cosh(2\kappa)-\cos p}.
\label{qkernel}
\end{equation}
Equation \eqref{Lieb} is often called the Lieb equation.

Having defined $\rho(p)$ and $\Lambda$ one can easily get the
ground-state expectation values of the integrals of motion, $\langle
I_m\rangle,$ in the limit \eqref{lattlim}. In particular,
\begin{equation}
\frac{\langle I_1\rangle}M=\chi^2 \int_{-\Lambda}^\Lambda dp\,
\rho(p)\cos p, \label{I1g}
\end{equation}
and
\begin{equation}
\frac{\langle I_2\rangle}M=\chi^2\left(1-\frac{\chi^2}2\right)
\int_{-\Lambda}^\Lambda dp\, \rho(p)\cos 2p. \label{I2g}
\end{equation}
The ground-state energy of the system can be obtained from
Eq.~\eqref{HviaI} or \eqref{EN}:
\begin{equation}
\frac{E_N}M=\frac{\langle H_q\rangle}M=
\frac4{\delta^2}\int_{-\Lambda}^\Lambda dp\,\rho(p)\sin^2\frac{p}2.
\end{equation}
We shall make use of the properties of $\langle I_m\rangle$ in
section \ref{sect:tqblmcl}.

\subsection{Hellmann-Feynman theorem \label{sect:HFi}}

In this section we use the Hellmann-Feynman theorem to derive an
identity for the ground-state average of some local operator in the
$q$-boson lattice model. This identity is given by
Eqs.~\eqref{HFTlocal} and \eqref{HFTidentity} below, and will be
used in section \ref{sect:result}.

It was shown in section \ref{sect:tqbm} that the eigenfunctions
\eqref{Bethestates} of the $q$-boson Hamiltonian \eqref{qHam} are
the eigenfunctions for the integrals of motion Eq.~\eqref{Imtilde}
and their Hermitian conjugate Eq.~\eqref{I-m}. Since
$I_1|\mathrm{gs}\rangle=I_1^\dagger|\mathrm{gs}\rangle,$
Eq.~\eqref{Imgs}, one has
\begin{equation}
\frac{d}{d q} \langle I_1 \rangle =\left\langle\frac{d}{d q} I_1
\right\rangle. \label{HFT}
\end{equation}
where $\langle\cdots\rangle$ denotes the ground state average and
$q$ is the deformation parameter defined by Eq.~\eqref{qalg0}.
Equation \eqref{HFT} is known under the name of the Hellmann-Feynman
theorem. The explicit expression for $I_1$ via local fields
$B_n^\dagger$ and $B_n$ is given by Eqs.~\eqref{IJ} and \eqref{J1}.
Using the translational invariance of the ground state, implying
that
\begin{equation}
\langle B^\dagger_{j_1}\dots B_{j_m}\rangle = \langle
B^\dagger_{j_1+k}\dots B_{j_m+k}\rangle \label{transinv}
\end{equation}
where $k$ is an arbitrary integer, equation \eqref{HFT} can be
written as follows:
\begin{equation}
\frac{d}{d q}\langle\chi^{-2}I_1\rangle=M \left\langle\frac{d}{dq}
(B^\dagger_{j}B_{j+1}) \right\rangle. \label{HFTlocal}
\end{equation}

The dependence of the operators $B_j$ and $B^\dagger_j$ on the
parameter $q$ can be analyzed using the representation
\eqref{Bn=bn}. One gets from this representation
\begin{align}
\frac{d}{dq} B_j = [N_j+1]^{-1/2}_q \frac{d}{dq} [N_j+1]^{1/2}_q
B_j,
\label{difBj2}\\
\frac{d}{dq} B_j^\dagger= B_j^\dagger [N_j+1]^{-1/2}_q \frac{d}{dq}
[N_j+1]^{1/2}_q. \label{difBj1}
\end{align}
Introducing the notation
\begin{equation}
g(q,x)= [x+1]_q^{-1/2}\frac{d}{dq} [x+1]_q^{1/2}=
\frac1q\left(\frac{x+1}{q^{2x+2}-1}-\frac{1}{q^2-1}\right)
\label{gdef}
\end{equation}
and combining Eqs.~\eqref{difBj2} and \eqref{difBj1} one finds
\begin{equation}
\frac{d}{dq} (B^\dagger_j B_{j+1})=B^\dagger_j g(q,N_j) B_{j+1}+
B^\dagger_j g(q,N_{j+1}) B_{j+1}. \label{HFTidentity}
\end{equation}

\section{Noether currents \label{sect:noether}}

Noether's theorem is an important ingredient of Quantum Field
Theory. In short, it states that symmetries imply conservation laws.
More precisely, if the action of a system is invariant under an
infinitesimal transformation of the fields, then there exists a
function of these fields whose divergence is zero. This function is
called the Noether current associated with the symmetry. One can say
more about this function if the symmetry transformation leaves the
Lagrangian (or, even better, the Lagrangian density) and not just
the action invariant \cite{W-95}.

For integrable models, however, it is often more natural to work
within the Hamiltonian formalism rather than within the Lagrangian
one. This is due to the fact that explicit expressions for the
integrals of motion are usually known in these models. In case of
the $q$-boson lattice model the local integrals of motion are
generated by Eq.~\eqref{Imtilde} and explicit expressions for
$\mathcal{J}_\tau^{(m)},$ like \eqref{J1}--\eqref{J3}, can be, in
principle, written down up to an arbitrarily large $m.$ The operator
$\mathcal{J}_\tau^{(m)}$ can be recognized as the imaginary-time
component of the conserved current $\mathcal{J}^{(m)}.$ By
calculating the commutator of $\mathcal{J}_\tau^{(m)}$ with the
Hamiltonian one gets the space component $\mathcal{J}_x^{(m)}$ of
the corresponding conserved current. The continuity equation for
$\mathcal{J}^{(m)}$ will be extensively used further derivations.

In section \ref{sect:NcHf} we recall briefly the notion of the
Noether currents in the Hamiltonian formalism. Noether currents in
the Lieb-Liniger model are considered in section \ref{sect:NCLL}. We
demonstrate the problem of making an unambiguous definition of
higher integrals of motion in this model. It is because of this
problem that we are working with the $q$-boson lattice
regularization (described in section \ref{sect:qblm}) of the
Lieb-Liniger model. Noether currents in the $q$-boson lattice model
are considered in section \ref{sect:ncqlm}. Finally, in section
\ref{sect:lgtqblm} we subject the $q$-boson lattice model to a local
gauge transformation and study the behavior of the Noether currents
under this transformation.

\subsection{Noether currents in the Hamiltonian formalism \label{sect:NcHf}}
To begin with, let us introduce some notation. It is often more
convenient in Quantum Field Theory to work with the Euclidean
(imaginary) time $\tau$ rather than with the Minkowski time $t:$
\begin{equation}
\tau=it. \label{ttau}
\end{equation}
We denote by $g_{\mu\nu}$ the metric tensor for two-dimensional
space-time theories\footnote{We choose $g_{\tau\tau}=-1$ and
$g_{xx}=1$ in Minkowski space-time.}:
\begin{equation}
g_{\mu\nu}=\left\{\begin{array}{ll}\mathrm{diag}(1,-1)&\qquad\text{Minkowski}\\
\mathrm{diag}(1,1)&\qquad\text{Euclidean}\end{array}\right.
\end{equation}
The summation is performed over the contracted indices,
\begin{equation}
a_\mu b^\mu\equiv \sum_\mu a_\mu b^\mu
\end{equation}
and the rules for converting between covariant and contravariant indices are
\begin{equation}
a_\mu=g_{\mu\nu}a^\nu, \qquad a^\mu=g^{\mu\nu}a_\nu, \qquad
g_{\mu\nu}g^{\nu\sigma}=\delta_\mu^\sigma.
\end{equation}

The Hamiltonian $H$ of a theory can be defined as the generator of
time-translations for an arbitrary operator $\mathcal{O}:$
\begin{equation}
\mathcal{O}(x,t)=e^{itH}\mathcal{O}(x,0)e^{-itH}. \label{OH}
\end{equation}
We assume that the Hamiltonian is time-independent:
\begin{equation}
\frac{\partial H}{\partial t}=\frac{dH}{dt}=0.
\end{equation}
Then the equation of motion for $\mathcal{O}(x,t)$ following from
Eq.~\eqref{OH} is
\begin{equation}
\frac{\partial}{\partial t}\mathcal{O}_M(x,t)=
\frac{d}{dt}\mathcal{O}_M(x,t)=i[H,\mathcal{O}_M(x,t)]. \label{evt}
\end{equation}
We note that all the dependence on $t$ of the operator
$\mathcal{O}_M(x,t)$ is explicitly given by the evolution operator
$e^{itH},$ Eq.~\eqref{OH}, and there is no need to distinguish the
partial $\partial/\partial t$ and full $d/dt$ derivatives. The
subscript $M$ is used in Eq.~\eqref{evt} to indicate that it is
written in the Minkowski time $t.$ Written in the Euclidean time
$\tau,$ Eq.~\eqref{ttau}, the equation of motion \eqref{evt} takes
the form
\begin{equation}
\frac{\partial}{\partial\tau}\mathcal{O}_E(x,\tau)=
\frac{d}{d\tau}\mathcal{O}_E(x,\tau)=[H,\mathcal{O}_E(x,\tau)],
\label{evtau}
\end{equation}
where $\mathcal{O}_E(x,\tau)=\mathcal{O}_M(x,t).$ We will mainly
work with the Euclidean time and therefore we drop the subscript $E$
to shorten notation. We shall also drop one or both of the arguments
$x,\tau$ in $\mathcal{O}(x,\tau)$ when they can be recovered from
the context.

Impose a {\it locality} condition onto the operator
$\mathcal{O}(x):$ suppose that this operator, taken at a point $x,$
depends on the basis fields $\psi$ and $\psi^\dagger$ (and their
derivatives) taken at this point exclusively. Equations \eqref{Nloc}
and \eqref{Ploc} provide us with examples of the operators of such
type. If, in addition, $\int_0^L dx\,\mathcal{O}(x)$ is an integral
of motion,
\begin{equation}
[H,\int_0^L dx\,\mathcal{O}(x)]=0, \label{commint}
\end{equation}
then
\begin{equation}
[H,\mathcal{O}(x)]=-\partial_x \mathcal{J}_x(x). \label{Jxdef}
\end{equation}
The symbol $\mathcal{J}_x$ is called the $x$-component of the
conserved (Noether) current. The operator $\mathcal{O}$ itself plays
a role of the $\tau$-component of the Noether current:
\begin{equation}
\mathcal{J}_\tau(x)=\mathcal{O}(x). \label{Jtaudef}
\end{equation}
Indeed, combining Eqs.~\eqref{evtau}, \eqref{Jxdef}, and
\eqref{Jtaudef} one gets
\begin{equation}
\partial_\tau\mathcal{J}_\tau+\partial_x\mathcal{J}_x\equiv
\partial_\tau\mathcal{J}^\tau+\partial_x\mathcal{J}^x=0 \qquad
(\text{Euclidean}) \label{cont}
\end{equation}
which is the continuity equation for the conserved (Noether) current
$(\mathcal{J}_\tau,\mathcal{J}_x).$ We thus derived Noether's
theorem within the Hamiltonian formalism: to every integral of
motion, Eq.~\eqref{commint}, there corresponds a conserved current,
Eq.~\eqref{cont}. It should be stressed that the time derivative
$\partial_\tau\mathcal{J}_\tau$ is defined by the right hand side of
Eq.~\eqref{evtau}; having $\partial_\tau\mathcal{J}_\tau,$ one can
use Eq.~\eqref{cont} to calculate $\mathcal{J}_x.$

\subsection{Noether currents in the Lieb-Liniger model \label{sect:NCLL}}

We now consider the Lieb-Liniger model \eqref{qContHam}. The local
density $\mathcal{N}(x)$ of the number operator Eq.~\eqref{Ndef} is
\begin{equation}
\mathcal{N}(x)\equiv\mathcal{J}^{(0)}_\tau(x)=\psi^\dagger(x)\psi(x). \label{Nloc}
\end{equation}
Commuting $\mathcal{N}(x)$ with the Hamiltonian \eqref{qContHam} one gets
\begin{equation}
[H,\mathcal{N}(x)]=-\partial_x \mathcal{J}^{(0)}_x(x),
\end{equation}
where
\begin{equation}
\mathcal{J}^{(0)}_x(x)= [\partial_x\psi^\dagger(x)]\psi(x)-
\psi^\dagger(x)\partial_x\psi(x). \label{J0xco1}
\end{equation}
Another conserved current is the current associated with the local
density $\mathcal{P}(x)$ of the momentum operator \eqref{Pdef}
\begin{equation}
\mathcal{P}(x)= \frac{i}2
\left\{\left[\partial_x\psi^\dagger(x)\right]\psi(x)-
\psi^\dagger(x)\partial_x\psi(x) \right\}. \label{Ploc}
\end{equation}
Commuting $\mathcal{P}(x)$ with the Hamiltonian \eqref{qContHam} one gets
\begin{equation}
[H,\mathcal{P}(x)]=-\partial_x \mathcal{M}(x) \qquad
(\text{ill-defined, see Eq.~\eqref{epm}}), \label{HPcomm}
\end{equation}
where
\begin{equation}
\mathcal{M}(x)=\frac{i}2\left\{
\left[\partial^2_x\psi^\dagger(x)\right]\psi(x)-
2\left[\partial_x\psi^\dagger(x)\right]\partial_x\psi(x)+
\psi^\dagger(x)\partial^2_x\psi(x)- 2c\psi^\dagger(x)^2\psi(x)^2
\right\}. \label{J1xLL}
\end{equation}

Let us discuss Eq.~\eqref{J1xLL} in more detail. In getting this
expression, one necessarily introduces objects like
$\psi^\dagger(x)\partial_x^3\psi(x)$ (this is clearly seen from
Eq.~\eqref{HPcomm}) and we have assumed that such objects are
well-defined. This is, however, an {\it incorrect} assumption. To
show this we use results from Ref.~\cite{OD-03}. Consider the ground
state average $\langle\psi^\dagger(x)\psi(0)\rangle.$ When $x\to0$
one can expand $\psi^\dagger(x)$ in the Taylor series:
\begin{multline}
\langle\psi^\dagger(x)\psi(0)\rangle= \langle\psi^\dagger(0)\psi(0)+
\frac{x}{1!}\partial_\epsilon\psi^\dagger(\epsilon)\psi(0) \\
+\left.\frac{x^2}{2!}\partial^2_\epsilon\psi^\dagger(\epsilon)\psi(0)+
\frac{x^3}{3!}\partial^3_\epsilon\psi^\dagger(\epsilon)\psi(0)+
\cdots\rangle\right|_{\epsilon=0}. \label{singexpansion}
\end{multline}
All the terms written explicitly on the right hand side of
Eq.~\eqref{singexpansion} can be found in Ref.~\cite{OD-03}:
\begin{equation}
\langle\psi^\dagger(0)\psi(0)\rangle=D,
\end{equation}
where $D$ is the average density, Eq.~\eqref{ddef},
\begin{equation}
\left.\langle\partial_\epsilon\psi^\dagger(\epsilon)
\psi(0)\rangle\right|_{\epsilon\to0}=0,
\end{equation}
and
\begin{equation}
\left.\langle\partial^2_\epsilon\psi^\dagger(\epsilon)
\psi(0)\rangle\right|_{\epsilon\to0}=\mathrm{const}(\gamma)\,D^3,
\end{equation}
where $\mathrm{const}(\gamma)$ depends on $\gamma,$
Eq.~\eqref{gamma}, exclusively. Finally,
\begin{equation}
\left.\langle\partial^3_\epsilon\psi^\dagger(\epsilon)
\psi(0)\rangle\right|_{\epsilon\to0}=
\mathrm{sgn}(\epsilon)\times\mathrm{const}(\gamma)\,D^4 \label{disc}
\end{equation}
where
\begin{equation}
\mathrm{sgn}(\epsilon)=
\left\{\begin{array}{rl}+1&\qquad\epsilon>0\\
-1&\qquad\epsilon<0\end{array}\right. .
\end{equation}

Equation \eqref{disc} is of crucial importance. It shows that
\begin{equation}
[\partial^3_x\psi^\dagger(x)]
\psi(x+\epsilon)\ne[\partial^3_x\psi^\dagger(x)] \psi(x-\epsilon)
\qquad\text{when}\qquad \epsilon\to0  \label{epm}
\end{equation}
so one should indicate explicitly how the point-splitting procedure
is performed whenever writing
$[\partial_x^3\psi^\dagger(x)]\psi(x)$. The same is true for other
operator products containing $\partial_x^3,$ (in particular,
Eqs.~\eqref{HPcomm} and \eqref{J1xLL} need such a prescription) and
more generally, for operator products containing the derivatives
$\partial_x^n$ with $n\ge3.$ We perform the point-splitting
procedure by putting a system on a lattice and we discuss the
corresponding Noether currents is section \ref{sect:ncqlm}.

\subsection{Noether currents in the $q$-boson lattice model \label{sect:ncqlm}}

It follows from the results of sections \ref{sect:imqbm} and
\ref{sect:NcHf} that the $q$-boson lattice model contains an
infinite hierarchy of Noether currents. Since this model is a
lattice model, all currents are well-defined. This is a major
advantage as compared to the Lieb-Liniger model, which, as it was
shown below Eq.~\eqref{J1xLL}, suffers from short-range
singularities. We obtain in the present section various relations
between conserved currents in the $q$-boson model. These relations
will be exploited in sections \ref{sect:bosonization} and
\ref{sect:stringid}.

The lattice version of the continuity equation \eqref{cont} for a
conserved current $\mathcal{J}=(\mathcal{J}_\tau, \mathcal{J}_x)$ is
\begin{equation}
\frac{\partial}{\partial \tau} \mathcal{J}_\tau(n, \tau) +
\frac{1}{\delta}\left[ \mathcal{J}_x(n, \tau)-\mathcal{J}_x(n-1,
\tau) \right] = 0 \label{latticecontinuity}
\end{equation}
where, according to Eqs.~\eqref{evtau} and \eqref{Jtaudef}, the
derivative of $\mathcal{J}_\tau$ with respect to $\tau$ is defined
as follows:
\begin{equation}
\frac{\partial}{\partial\tau} \mathcal{J}_\tau \equiv [H_q,
\mathcal{J}_\tau]. \label{dtau}
\end{equation}
The Hamiltonian \eqref{qHam} commutes with the total number of
particles, Eq.~\eqref{N}. The $\tau$-component,
$\mathcal{J}_\tau^{(0)}(n),$ of the corresponding local current is
given by Eq.~\eqref{NJ0}:
\begin{equation}
\mathcal{J}^{(0)}_\tau(n) =\frac{1}{\delta} N_n. \label{chden}
\end{equation}
We calculate the commutator $[H_q,N_n]$ with the help of
Eq.~\eqref{qalg}, then substitute the resulting expression into
Eq.~\eqref{latticecontinuity} and get
\begin{equation}
\mathcal{J}^{(0)}_x(n) = \frac{1}{\delta^2}(B^\dagger_{n+1}B_n -
B^{\dagger}_n B_{n+1}). \label{curden}
\end{equation}
Using Eq.~\eqref{J1} one can write Eq.~\eqref{curden} as follows:
\begin{equation}
\mathcal{J}^{(0)}_x(n)= -\frac1{\delta\chi^2}
\left[\mathcal{J}^{(1)}_\tau(n)-\mathcal{J}^{(-1)}_\tau(n)\right].
\label{Jx0Jtau}
\end{equation}
In the continuum limit \eqref{qCont} equations \eqref{chden} and
\eqref{curden} become
\begin{align}
&\mathcal{J}^{(0)}_\tau(x)= \psi^\dagger(x)\psi(x), \label{J0tauco2}\\
&\mathcal{J}^{(0)}_x(x)=
\left[\partial_x\psi^\dagger(x)\right]\psi(x)-
\psi^\dagger(x)\partial_x\psi(x), \label{J0xco2}
\end{align}
thus reproducing the expressions \eqref{Nloc} and \eqref{J0xco1} for
the $\tau$- and $x$-components of the conserved current
$\mathcal{J}^{(0)}$ in the Lieb-Liniger model.

The $x$-component of the current $\mathcal{J}^{(1)}(n)$ can be
calculated in very much the same way as $\mathcal{J}_x^{(0)}(n).$
Indeed, the $\tau$-component of $\mathcal{J}^{(1)}(n)$ is given by
Eq.~\eqref{J1}. Substituting this expression into
Eq.~\eqref{latticecontinuity} and calculating the commutator
$[H_q,\mathcal{J}_\tau^{(1)}(n)]$ with the help of
Eqs.~\eqref{qalg0}, \eqref{qalg}, and \eqref{BBalt}, one gets
\begin{equation}
\mathcal{J}_x^{(1)}(n)= \frac{\chi^2}{\delta^2}\left[-B^\dagger_n
B_{n+2}+ \chi^2B^\dagger_nB^\dagger_{n+1}B_{n+1}B_{n+2}+ B^\dagger_n
B_n\right], \label{Jx1}
\end{equation}
where $\chi$ is defined by Eq.~\eqref{chi}. For the $x$-component of
the current $\mathcal{J}^{(-1)}(n)$ one gets, taking into account
the relation \eqref{J-m},
\begin{equation}
\mathcal{J}_x^{(-1)}(n)=-\left[\mathcal{J}_x^{(1)}(n)\right]^\dagger.
\label{Jx-1}
\end{equation}

Having Eqs.~\eqref{Jx1} and \eqref{Jx-1} one can easily calculate
the operator $\mathcal{M},$ Eq.~\eqref{HPcomm}, in the $q$-boson
lattice model. The operator $\mathcal{M}$ is defined unambiguously
on the lattice by equation \eqref{latticecontinuity}:
\begin{equation}
[H_q,\mathcal{P}(n)]=-\frac1\delta[\mathcal{M}(n)-\mathcal{M}(n-1)],
\label{lattPM}
\end{equation}
where $\mathcal{P}(n)$ is the density of the momentum operator
\eqref{Pqdef}:
\begin{equation}
\mathcal{P}(n)=-\frac{i}2\frac1{\delta^2}(B^\dagger_nB_{n+1}-
B^\dagger_{n+1}B_n)=
-\frac{i}2\frac1{\delta\chi^2}\left[\mathcal{J}_\tau^{(1)}(n)-
\mathcal{J}_\tau^{(-1)}(n)\right]
\end{equation}
(recall that the local densities on a lattice are defined according
to Eq.~\eqref{IJ}). The resulting expression for $\mathcal{M}(n)$ is
\begin{equation}
\mathcal{M}(n)=-\frac{i}2\frac1{\delta\chi^2}
\left[\mathcal{J}_x^{(1)}(n)- \mathcal{J}_x^{(-1)}(n)\right].
\label{Mn}
\end{equation}
In the continuum limit \eqref{qCont} this expression transforms to
Eq.~\eqref{J1xLL}.

Finally, we note the following property of $\mathcal{J}_x^{(m)}(n)$
\begin{equation}
[N,\mathcal{J}_x^{(m)}(n)]=0, \qquad m=0,\pm1,\pm2,\dots.
\label{NJxcomm}
\end{equation}
The proof of this expression is the same as that of
Eq.~\eqref{NJtaucomm}. Equations \eqref{NJtaucomm} and
\eqref{NJxcomm} play an important role in the bosonization procedure
for the $q$-boson lattice model, carried out in section
\ref{sect:bnc}.

\subsection{Local gauge transformations in the $q$-boson lattice model \label{sect:lgtqblm}}

The operator $N$ in the $q$-boson model generates the global $U(1)$
rotation (it is often called the global $U(1)$ gauge transformation)
of the fields $B_n$ and $B^\dagger_n$:
\begin{equation}
e^{-i\phi N} B_n e^{i \phi N} = B_n e^{i\phi}, \qquad e^{-i\phi N}
B^\dagger_n e^{i \phi N} = B^\dagger_n e^{-i\phi}.
\label{U(1)rotation}
\end{equation}
The $\tau$-component of the Noether current $\mathcal{J}^{(m)}(n)$
is invariant under this rotation:
\begin{equation}
e^{-i\phi N}\mathcal{J}^{(m)}_\tau(n)e^{i \phi N}=
\mathcal{J}^{(m)}_\tau(n), \qquad m=0,\pm1,\pm2,\ldots .
\end{equation}
Whenever one works with a system possessing a global symmetry
transformation, it is useful to extend this transformation to a {\it
local} one. The local gauge transformation coming from the extension
of the global $U(1)$ symmetry is performed by a unitary operator
\begin{equation}
Q_\epsilon=\exp\left[i \epsilon \delta \sum_{n=1}^{M}
\eta(n)\mathcal{J}_\tau^{(0)}(n)\right], \label{boost}
\end{equation}
where $\mathcal{J}_\tau^{(0)}(n)$ is the density operator
\eqref{chden}, the parameter $\epsilon$ is an arbitrary real number,
and $\eta(n)$ is an arbitrary function of the lattice coordinate
$n.$ The $q$-boson fields $B_n$ and $B^\dagger_n$ (called ``matter
fields'' in Quantum Field Theory) are transformed by the operator
\eqref{boost} as follows
\begin{equation}
Q^{-1}_\epsilon B_n Q_\epsilon= e^{i \epsilon \eta(n)} B_n, \qquad
Q^{-1}_\epsilon B_n^{\dagger} Q_\epsilon = e^{-i \epsilon \eta(n)}
B_n^\dagger. \label{BBdboost}
\end{equation}
For $\eta(n)=\mathrm{const},$ Eq.~\eqref{BBdboost} reproduces the
transformation law \eqref{U(1)rotation}.

The $\tau$-component of $\mathcal{J}^{(m)}(n)$ is not invariant
under the action of $Q_\epsilon;$ its evolution is described by the
operator
\begin{equation}
\mathcal{J}_\tau^{(m)}(n,\epsilon)\equiv Q^{-1}_\epsilon
\mathcal{J}_\tau^{(m)}(n) Q_\epsilon, \qquad m=0,\pm1,\pm2,\ldots .
\label{Jne}
\end{equation}
For the case when the function $\eta(n)$ is a linear function of
$n,$
\begin{equation}
\eta(n)=a\delta n+b \label{etalin}
\end{equation}
one can get the following representation of the right hand side of
Eq.~\eqref{Jne}:
\begin{equation}
\mathcal{J}_\tau^{(m)}(n,\epsilon)= e^{i\epsilon a\delta m}
\mathcal{J}_\tau^{(m)}(n), \qquad m=0,\pm1,\pm2,\dots.
\label{JBoost}
\end{equation}
For $\mathcal{J}_\tau^{(1)},$ $\mathcal{J}_\tau^{(2)},$ and
$\mathcal{J}_\tau^{(3)}$ this expression can be checked by applying
$Q_\epsilon$ to Eqs.~\eqref{J1}, \eqref{J2}, and \eqref{J3},
respectively. By investigating the structure of Eq.~\eqref{Imtilde}
one generalizes this calculation to the case of an arbitrary $m.$ It
follows from Eq.~\eqref{JBoost} that
\begin{equation}
\left.\frac{d}{d\epsilon}
\mathcal{J}_\tau^{(m)}(n,\epsilon)\right|_{\epsilon=0} = ia\delta m
\mathcal{J}_\tau^{(m)}(n). \label{JBinf}
\end{equation}
We shall need the ground state expectation value of this equation
\begin{equation}
\left.\left\langle\frac{d}{d\epsilon}
\mathcal{J}_\tau^{(m)}(n,\epsilon)\right\rangle\right|_{\epsilon=0}
= i a\delta m\langle \mathcal{J}_\tau^{(m)}\rangle.
\label{boostinfc}
\end{equation}

We will also need to know the action of $Q_\epsilon$ onto the
$x$-component of the current $\mathcal{J}^{(0)}(n)$,
Eq.~\eqref{curden}. Assuming that $\eta(n)$ is the linear function
of $n,$ Eq.~\eqref{etalin}, one gets from Eqs.~\eqref{Jx0Jtau} and
\eqref{JBinf}
\begin{equation}
\left.\frac{d}{d\epsilon}
\mathcal{J}_x^{(0)}(n,\epsilon)\right|_{\epsilon=0}=
-\frac{ia}{\delta \chi^2}
\left[\mathcal{J}_\tau^{(1)}(n)+\mathcal{J}_\tau^{(-1)}(n)\right]
\label{boosel}.
\end{equation}
Taking the ground state average and using Eqs.~\eqref{Imtrans} and
\eqref{Jmgs} one arrives at
\begin{equation}
\left.\left\langle\frac{d}{d\epsilon}
\mathcal{J}_x^{(0)}(n,\epsilon)\right\rangle\right|_{\epsilon=0} =
-\frac{2 ia}{\delta \chi^2 } \langle \mathcal{J}_\tau^{(1)}\rangle=
-\frac{2 ia}{\delta^2 \chi^2 } \frac{\langle I_1\rangle}M.
\label{boostel}
\end{equation}

We define now the currents $j^{(m)}$ by the formula
\begin{equation}
j^{(m)}_\mu \equiv \mathcal{J}^{(m)}_\mu -\langle
\mathcal{J}^{(m)}_\mu \rangle, \qquad \mu=\tau,x, \qquad
m=0,\pm1,\pm2,\ldots, \label{nocur}
\end{equation}
The continuity equation for $j^{(m)}$ is
\begin{equation}
\partial^\mu j_{\mu}^{(m)}=0.
\label{Noether1}
\end{equation}
The $\epsilon$-dependent current $j_\tau^{(m)}(n,\epsilon)$ is
defined in the same manner as $\mathcal{J}_\tau^{(m)}(n,\epsilon):$
\begin{equation}
j_\tau^{(m)}(n,\epsilon)\equiv Q^{-1}_\epsilon j_\tau^{(m)}(n)
Q_\epsilon, \qquad m=0,\pm1,\pm2,\ldots .
\end{equation}
It is obvious that
\begin{equation}
\left.\frac{d}{d\epsilon}j_\tau^{(m)}{(n,\epsilon)}\right|_{\epsilon=0}=
\left.\frac{d}{d\epsilon}\mathcal{J}_\tau^{(m)}{(n,\epsilon)}\right|_{\epsilon=0}.
\end{equation}
One can, therefore, write Eq.~\eqref{JBinf} as follows
\begin{equation}
\left.\frac{d}{d\epsilon}j_\tau^{(m)}{(n,\epsilon)}\right|_{\epsilon=0}=
ia\delta m \mathcal{J}_\tau^{(m)}(n), \qquad m=0,\pm1,\pm2,\dots .
\label{boostinf}
\end{equation}
In particular,
\begin{equation}
\left.\frac{d}{d\epsilon}
j_\tau^{(0)}(n,\epsilon)\right|_{\epsilon=0} =0. \label{notaue}
\end{equation}

\section{Bosonization of the $q$-boson lattice model. \label{sect:bosonization}}

In this section we discuss an effective field theory describing the
low-energy properties of the $q$-boson lattice model. This effective
field theory is the free boson theory in one space and one time
dimension, studied in great detail in many review articles and
textbooks, for example in Refs.~\cite{Gia-04,GNT-93}. Some of its
basic properties are reviewed briefly within the coordinate space
formulation of section \ref{sect:nnc}, the others are given within
the momentum state formulation of section \ref{sect:fbtms}. All the
correlation functions of the free boson theory can be calculated
explicitly, making it possible to classify the operators of the
theory according to their anomalous dimensions. This is the subject
of section \ref{sect:cfad}. In section \ref{sect:bnc} we perform the
so-called bosonization procedure: we establish the correspondence
between the operators of the microscopic theory (the $q$-boson
lattice model) in the low energy limit, and the operators of the
free boson theory (shortly, we take the bosonized limit of the
microscopic operators). We continue the bosonization procedure in
section \ref{sect:anc} where we express the $c$-number coefficients
in the bosonized representation of microscopic operators via
$\langle I_m\rangle,$ studied in section \ref{sect:qblm}.

\subsection{Free boson theory in coordinate space \label{sect:nnc}}

In the present section we review briefly the free boson theory in
the coordinate space formulation. The Lagrangian density in
Minkovski space is
\begin{equation}
\mathcal{L}_M=-\frac1{2\pi
K}\left[-\frac1v(\partial_t\phi)^2+v(\partial_x\phi)^2\right].
\label{LM}
\end{equation}
and the corresponding Lagrangian density in the Euclidean space is
\begin{equation}
\mathcal{L}_E=\frac1{2\pi
K}\left[\frac1v(\partial_\tau\phi)^2+v(\partial_x\phi)^2\right].
\label{LE}
\end{equation}
(the rules for converting between Minkovski and Euclidean spaces are
given in section \ref{sect:NcHf}). The parameter $v$ has the
dimension of velocity, the parameter $K$ is dimensionless, it is
often called Luttinger parameter. The action $S_M$ in Minkovski
space is defined as
\begin{equation}
S_M=\int dt\,dx\, \mathcal{L}_M.
\end{equation}
The model is placed on a ring of circumference $L,$ the boundary
conditions are discussed in the paragraph below Eq.~\eqref{Tord}. To
define the Euclidean action $S_E$ note that the weight function
$e^{iS_M},$ which appears in the functional integral formulation of
the theory, oscillates in Minkovski space, while in Euclidean space
it should decay rapidly and is conventionally written as $e^{-S_E}$.
Therefore
\begin{equation}
e^{-S_E}=e^{iS_M}
\end{equation}
and
\begin{equation}
S_E=\frac{1}{2\pi K}\int d\tau\,dx\,
\left[\frac{1}{v}(\partial_\tau\phi)^2+ v(\partial_x\phi)^2\right].
\label{SLut}
\end{equation}
The field $\phi$ is a free massless real scalar boson field. An
equation of motion for this field is the wave equation:
\begin{equation}
\partial_\tau^2\phi+ v^2\partial_x^2\phi=0. \label{eqm}
\end{equation}
The canonical momentum $\Pi$ (the field canonically conjugated to
$\phi$) is
\begin{equation}
\Pi=\frac{\partial\mathcal{L}_M}{\partial\partial_t\phi}= \frac1{\pi
Kv}\partial_t\phi=\frac{i}{\pi Kv}\partial_\tau\phi.
\label{Pitauphi}
\end{equation}
It is often convenient to work with the field $\theta(x),$ defined
as follows:
\begin{equation}
\Pi(x)=\frac1\pi\partial_x\theta(x). \label{Pitheta}
\end{equation}
We shall switch freely between $\partial_\tau\phi,$
$\partial_x\theta,$ and $\Pi$ in subsequent formulas. Being
quantized, the fields $\phi$ and $\Pi$ obey canonical equal-time
commutation relations
\begin{equation}
[\Pi(x),\phi(y)]=-i\delta(x-y). \label{canonical}
\end{equation}
The Hamiltonian $H$ of the system has the form
\begin{equation}
H=\int dx\,\mathcal{H}(x), \qquad
\mathcal{H}(x)=\frac{v}{2\pi}:K\left[\pi\Pi(x)\right]^2+
\frac1K\left[\partial_x\phi(x)\right]^2:,  \label{Hfb}
\end{equation}
where the symbol $::$ stands for normal ordering, discussed in
section \ref{sect:fbtms}. Another important ordering prescription is
the time-ordering $T.$ For any two boson operators $A(\tau)$ and
$B(\tau^\prime)$ it is defined as follows
\begin{equation}
T A(\tau)B(\tau^\prime)=\left\{\begin{array}{ll}
A(\tau)B(\tau^\prime),&\quad\tau>\tau^\prime\\
B(\tau^\prime)A(\tau),&\quad\tau<\tau^\prime\end{array} \right. .
\label{Tord}
\end{equation}
We assume that $T$ acts on all the operators standing on the right.

Boundary conditions are an important issue in the theory. We impose
periodic boundary conditions on the operators $\partial_x\phi$ and
$\partial_x\theta:$
\begin{equation}
\partial_x\phi(x+L)=\partial_x\phi(x), \qquad
\partial_x\theta(x+L)=\partial_x\theta(x), \label{dpdtb}
\end{equation}
Since the Hamiltonian \eqref{Hfb} contains $\partial_x\phi$ and
$\partial_x\theta,$ rather than these fields themselves, one has
\begin{equation}
\phi(x+L)=\phi(x)+\alpha_1, \qquad \theta(x+L)=\theta(x)+\alpha_2,
\label{ptb}
\end{equation}
where $\alpha_1$ and $\alpha_2$ are some operators. Their properties
are discussed in section \ref{sect:fbtms}.

\subsection{Free boson theory in momentum space \label{sect:fbtms}}

In this section we give the momentum space representation of the
free boson theory \eqref{Hfb}. This representation is a convenient
starting point for calculating the correlation functions of the
theory, like those considered in section \ref{sect:cfad}.

The momentum space representation of the $\phi$ and $\theta$ fields
is
\begin{multline}
\phi(x,t)=\phi_0+\pi vtK\frac{J}L-\pi
x\frac{N-N_0}{L}\\-\frac{i}{2}\sum_{q\ne0} \left|\frac{2\pi
K}{qL}\right|^{1/2} \mathrm{sgn}(q) e^{-iqx} \left(b^{\dagger}_q
e^{ivt|q|}+ b_{-q}e^{-ivt|q|}\right) \label{phi->bt}
\end{multline}
and
\begin{equation}
\theta(x,t)=\theta_0-\pi vt \frac{N-N_0}{KL}+\pi x\frac{J}{L}+
\frac{i}{2}\sum_{q\ne0} \left|\frac{2\pi}{qLK}\right|^{1/2} e^{-iqx}
\left(b^\dagger_q e^{ivt|q|} - b_{-q}e^{-ivt|q|}\right),
\label{theta->bt}
\end{equation}
respectively. The momentum space representation of the Hamiltonian
\eqref{Hfb} is:
\begin{equation}
H=\frac{\pi v}{2L}\left[\frac1K(N-N_0)^2+K J^2\right]+
v\sum_{q\ne0}|q|b^\dagger_q b_q. \label{luttp}
\end{equation}
The summation index $q$ in Eqs.~\eqref{phi->bt}--\eqref{luttp} runs
through the following set of values:
\begin{equation}
q=\frac{2\pi}Lj, \qquad j=\pm1,\pm2,\ldots.
\end{equation}
The operators $b^\dagger_q$ ($b_q$) are boson creation
(annihilation) operators obeying canonical commutation relations
\begin{equation}
[b_q,b^\dagger_{q^\prime}]=\delta_{qq^\prime}. \label{bbcomm}
\end{equation}
Thus, the last term in the right hand side of Eq.~\eqref{luttp}
represents a set of decoupled harmonic oscillators with the
frequencies $v|q|$. The first three terms on the right hand side of
Eqs.~\eqref{phi->bt} and \eqref{theta->bt} give the so-called
``zero-mode contribution''. They all commute with $b^\dagger_q$ and
$b_q;$ the only nontrivial commutation relations between themselves
are
\begin{equation}
[J,\phi_0]=-i, \qquad [N,\theta_0]=i. \label{zmcomm}
\end{equation}
Comparing Eq.~\eqref{ptb} with Eqs.~\eqref{phi->bt} and
\eqref{theta->bt}, one can easily express $\alpha_1$ and $\alpha_2$
via $J$ and $N-N_0$ ($N_0$ is a $c$-number). The normal ordering
symbol $::$ standing in Eqs.~\eqref{Hfb} and \eqref{vdef} means that
in any given monomial one should place the creation operators
$b_q^\dagger$ to the left of the annihilation operators $b_q.$


We denote the ground state of the theory as $|0\rrangle;$ the ground
state expectation value $\llangle\mathcal{O}\rrangle$ of an
arbitrary operator $\mathcal{O}$ is
$\llangle0|\mathcal{O}|0\rrangle.$ We use the symbol
$\llangle\cdots\rrangle$ for the ground state expectation value of
the free boson theory in order to distinguish it from the ground
state expectation value $\langle\cdots\rangle$ of the microscopic
model, Eq.~\eqref{g3def}. One has
\begin{equation}
b_k|0\rrangle=0 \qquad \text{for} \qquad k=\pm1,\pm2,\ldots
\label{vaccond}
\end{equation}
and
\begin{equation}
J|0\rrangle=0, \qquad N|0\rrangle=N_0|0\rrangle.
\end{equation}
Finally, we define the action of the operators $\phi_0$ and
$\theta_0$ on $|0\rrangle.$ We do this taking the exponent of these
operators\footnote{One can easily identify the operators $\phi_0$
and $\theta_0$ with the phase operators, canonically conjugated to
the number operators $N$ and $J.$ The phase operators are not
well-defined in the whole Hilbert space of the theory, while their
exponents are. Thus, the commutation relations \eqref{zmcomm} should
be understood as applied to the operators $e^{im\phi_0}$ and
$e^{in\theta_0}$ exclusively.}: the states $e^{im\phi_0}|0\rrangle$
and $e^{in\theta_0}|0\rrangle$ are non-vanishing states orthogonal
to the vacuum state:
\begin{equation}
\llangle e^{im\phi_0}\rrangle=\delta_{m0}, \qquad \llangle
e^{in\theta_0}\rrangle=\delta_{n0}, \label{wa}
\end{equation}
where $\delta_{m_1m_2}$ is the Kronecker $\delta$ symbol
\begin{equation}
\delta_{m_1m_2}=\left\{\begin{array}{ll}
1&m_1=m_2\\
0&m_1\ne m_2\end{array} \right. .
\end{equation}
Any operator within the free boson theory we shall work with, will
contain the fields $\phi_0$ and $\theta_0$ in the form
$e^{im\phi_0}$ and $e^{in\theta_0}$ exclusively.

\subsection{Correlation functions and spectrum on the anomalous dimensions in the free boson theory \label{sect:cfad}}
The correlation functions of the free boson theory, Eq.~\eqref{Hfb},
are known explicitly and are given in many textbooks on
one-dimensional physics~\cite{Gia-04,GNT-93}. They can be
calculated, for example, starting from the momentum space
representation, discussed in section~\ref{sect:fbtms}. The knowledge
of the correlation functions provides us with an efficient tool for
classifying the operator content of the theory. We classify the
operators according to their anomalous dimensions. The operators
with the lowest anomalous dimensions are the most relevant in the
low-energy sector of the microscopic model, assuming that this
sector can be mapped onto the free boson theory.

We start by considering the following correlation function of the
free boson theory
\begin{equation}
\llangle T \phi(x,\tau)\phi(x^\prime,\tau^\prime)\rrangle=-
\frac{K}{4}\ln[(x-x^\prime)^2+v^2 (\tau-\tau^\prime)^2].
\label{Gbose}
\end{equation}
Equation~\eqref{Gbose} is written assuming that the limit
$L\to\infty$ is taken. Moreover, to remove the ill-defined operator
$\phi_0$ from the fields $\phi(x,\tau)$ and
$\phi(x^\prime,\tau^\prime)$ one should differentiate them at least
once with respect to $x$ or $\tau$ and $x^\prime$ or $\tau^\prime.$
Thus, an example of a well-defined correlation function is
\begin{equation}
\llangle T
\partial_x\phi(x,\tau)\partial_{x^\prime}\phi(x^\prime,\tau^\prime)\rrangle=
-\frac{K}2\frac{(x-x^\prime)^2-v^2(\tau-\tau^\prime)^2}
{[(x-x^\prime)^2+v^2(\tau-\tau^\prime)^2]^2}. \label{Gppxx}
\end{equation}
One can see that the correlation function \eqref{Gppxx} exhibits the
power-law decay with the exponent equal to two as $x-x^\prime$ and
$\tau-\tau^\prime$ goes to infinity. This exponent defines the
so-called {\it anomalous dimension} (or {\it conformal dimension})
of the operator $\partial_x\phi:$ it is equal to one. Calculating
the correlation function $\llangle
T\partial_x\theta\partial_{x^\prime}\theta\rrangle$ one gets that
the anomalous dimension of the operator $\partial_x\theta$ is equal
to one as well. The generalization of these results is obvious: the
anomalous dimension of the operator
$\partial^n_\tau\partial^m_x\phi$ is equal to $n+m.$ We recall that
$\partial_x\theta\sim\partial_\tau\phi,$ which follows from
Eqs.~\eqref{Pitauphi} and \eqref{Pitheta}.

Next, we define two one-parametric families, $\mathcal{V}_{m}(x)$
and $\mathcal{W}_{m}(x),$ of the so-called {\it vertex operators},
\begin{equation}
\mathcal{V}_{m}(x)=:e^{im\phi(x)}:, \qquad
\mathcal{W}_{m}(x)=:e^{im\theta(x)}:, \label{vdef}
\end{equation}
where $m$ is an arbitrary real number. One has
\begin{align}
&\llangle
T\mathcal{V}_{m_1}(x,\tau)\mathcal{V}_{-m_2}(x^\prime,\tau^\prime)
\rrangle = \frac{\delta_{m_1m_2}}{[(x-x^\prime)^2+
v^2(\tau-\tau^\prime)^2]^{Km_1^2/4}}, \label{vcorr}\\
&\llangle
T\mathcal{W}_{m_1}(x,\tau)\mathcal{W}_{-m_2}(x^\prime,\tau^\prime)
\rrangle = \frac{\delta_{m_1m_2}}{[(x-x^\prime)^2+
v^2(\tau-\tau^\prime)^2]^{m_1^2/4K}}, \label{wcorr}\\
&\llangle T\mathcal{V}_{m_1}(x,\tau)
\mathcal{W}_{-m_2}(x^\prime,\tau^\prime)\rrangle=0.
\end{align}
Thus, the anomalous dimensions of the operators $\mathcal{V}_m$ and
$\mathcal{W}_m$ are equal to $Km^2/4$ and $m^2/4K,$ respectively.
There are no constraints within the free boson theory on the
possible values of $K$ and $m.$ However, when one uses the free
boson theory to describe the low-energy sector of a microscopic
theory, some constraints can appear. In particular, to describe the
low-energy sector of the Lieb-Liniger (as well as $q$-boson) model,
the possible values of $m$ in \eqref{vdef} should be restricted by
the following discrete set:
\begin{equation}
m=\pm1, \pm2,\ldots .
\end{equation}
This condition will become clear from the arguments of sections
\ref{sect:bnc} and \ref{sect:gfcc}.

We are interested in the following correlation function of an
operator $A(x,\tau)$
\begin{equation}
\llangle TA(x,\tau)A(x^\prime,\tau^\prime)\rrangle \qquad \text{as}
\qquad x-x^\prime\to\infty \qquad \text{and} \qquad
\tau-\tau^\prime\to\infty.
\end{equation}
To calculate this function, we assume that the operator $A=A(\phi)$
can be expanded in a series and that every term of this series is a
$\phi$-dependent operator with some conformal dimension. The fields
with the lowest conformal dimension are either $\partial_x\phi$ and
$\partial_\tau\phi$ with the conformal dimension $1$ or the vertex
operators \eqref{vdef} with the conformal dimensions $\dim
\mathcal{V}_m=Km^2/4$ and $\dim \mathcal{W}_m=m^2/4K.$ Which of
these four operators has (or have) the lowest dimension, depends on
the values of $K$ and $m.$ Thus, we write
\begin{equation}
A(\phi)= a_1\partial_x\phi+a_2\partial_\tau\phi+ a_3\mathcal{V}_{m_1}+
a_4\mathcal{W}_{m_2}+\text{h.o.t.}, \label{Aphi}
\end{equation}
where the symbol ``h.o.t'' stands for the subleading terms, and
$a_1,\ldots,a_4$ are some $c$-number coefficients. We want to stress
that the correct usage of the expansion \eqref{Aphi} is as follows:
among four terms written explicitly on the right hand side, one
should select the one (ones) with the lowest conformal dimension,
and the rest should be included into the ``h.o.t.''. In other words,
if, for instance, $\mathcal{V}_{m_1}$ has the lowest conformal
dimension, the first subleading term does not necessarily come from
the remaining three terms written explicitly on the right hand side
of Eq.~\eqref{Aphi}.

\subsection{Bosonization of the $q$-boson lattice model: operator relations  \label{sect:bnc}}

In this section we shall exploit an assumption that the low-energy
physics of the $q$-boson lattice model, Eq.~\eqref{qHam}, is
described by the Hamiltonian \eqref{Hfb}. In other words, we will
exploit an assumption that the $q$-boson lattice model (and its
continuum limit, the Lieb-Liniger model) belongs to a universality
class usually referred as the Luttinger Liquid \cite{Gia-04,GNT-93}.
This assumption becomes useful in practice when the correspondence
between the operators of the microscopic theory (the fields $B_n$
and $B^\dagger_n$ in our case) and the operators of the free boson
theory is established (shortly, establishing this correspondence,
one ``bosonizes'' the microscopic theory).

Our aim is to bosonize the currents $j^{(m)},$ Eq.~\eqref{nocur}. We
write them in the form given by Eq.~\eqref{Aphi} and we should find
the spectrum of the anomalous dimensions and the values of
$a_1,\dots,a_4$ from the properties of the microscopic theory. To
distinguish the microscopic currents $j^{(m)}$ from their bosonized
form, we write the latter as $\mathbf{j}^{(m)}$:
\begin{align*}
&j^{(m)}&&\text{ the current in the microscopic ($q$-boson) theory}\\
&\mathbf{j}^{(m)}&&\text{ an operator within the free boson theory
corresponding to }j^{(m)}
\end{align*}

(i) We associate the operator $N$ appearing in section
\ref{sect:fbtms} with the particle number operator of the
microscopic model. The particle number operator commutes with
$j^{(m)},$ Eqs.~\eqref{NJtaucomm}, \eqref{NJxcomm}, and
\eqref{nocur}, while it follows from Eqs.~\eqref{theta->bt},
\eqref{zmcomm}, and \eqref{vdef} that $[N,\mathcal{W}_m]\ne0.$
Therefore, the vertex operators $\mathcal{W}_m$ are not present in
the expansion \eqref{Aphi} of the currents $j^{(m)}.$

(ii) We require the bosonized form of $j^{(m)}$ to be unchanged
under the transformation $x\to x+L.$ It can be easily seen from
Eqs.~\eqref{phi->bt} and \eqref{vdef} that for the vertex operators
$\mathcal{V}_m$ this implies the following constraints on the
possible values of $m:$
\begin{equation}
m=0,\pm2,\pm4,\dots
\end{equation}
Upon bosonization, the ground state average $\langle\cdots\rangle$
should be replaced with the average $\llangle\cdots\rrangle.$ Since
$\langle j^{(m)}\rangle=0,$ one has $\llangle
\mathbf{j}^{(m)}\rrangle=0$ in the bosonized theory and the operator
$\mathcal{V}_{m=0}=1$ is not present in the expansion \eqref{Aphi}
of $j^{(m)}.$

(iii) One can see from (II) and Eq.~\eqref{vcorr} that the lowest
possible anomalous dimension of $\mathcal{V}_m$ is $K.$ It can be
easily shown, using the results of Ref.~\cite{H2-81}, that
\begin{equation}
K>1 \qquad \text{as} \qquad \gamma>0 \label{K>1}
\end{equation}
in the Lieb-Liniger model. For the $q$-boson lattice model, we shall
always assume that we are sufficiently close to the continuum limit
to satisfy Eq.~\eqref{K>1} as well. Taking Eq.~\eqref{K>1} into
account, one bosonizes $j^{(m)}$ as follows
\begin{equation}
\mathbf{j}^{(m)}_\tau  = -\frac{\alpha_{m}}\pi\partial_x\phi+
\frac{\beta_{m}} {\pi} \partial_\tau\phi+ \text{h.o.t.}
\label{jmtau}
\end{equation}
and
\begin{equation}
\mathbf{j}^{(m)}_x = \frac{\alpha_{m}}\pi\partial_\tau\phi+
\frac{\beta_{m} v^2}\pi\partial_x\phi+ \text{h.o.t.} \label{jmx}
\end{equation}
The bosonized form of the continuity equation \eqref{Noether1},
\begin{equation}
\partial^\mu \mathbf{j}_\mu^{(m)}=0
\end{equation}
together with the equation of motion \eqref{eqm} imply that the
coefficients $\alpha_m$ and $\beta_m$ in Eq.~\eqref{jmtau} are the
same as in Eq.~\eqref{jmx}.

(iv) The coefficients $\alpha_0$ and $\beta_0$ are
\begin{equation}
\alpha_0=1, \qquad \beta_0=0, \label{a1b0}
\end{equation}
and therefore
\begin{align}
&\mathbf{j}^{(0)}_\tau = -\frac{1}{\pi}\partial_x\phi+\text{h.o.t.} \label{jet}\\
&\mathbf{j}^{(0)}_x = \frac{1}{\pi}\partial_\tau\phi+\text{h.o.t.}
\label{jex}
\end{align}
The proof of Eq.~\eqref{a1b0} will be given in section
\ref{sect:gfcc}.

(v) It follows from (iv) that $\mathbf{j}_\tau^{(m)}$,
Eq.~\eqref{jmtau}, can be written in terms of the components of the
current $\mathbf{j}^{(0)}$:
\begin{equation}
\mathbf{j}^{(m)}_\tau  = \alpha_m \mathbf{j}_\tau^{(0)}+\beta_m
\mathbf{j}_x^{(0)} +\text{h.o.t.}, \qquad m=0,\pm1,\pm2,\dots.
\label{jjel}
\end{equation}
To use this relation, we need to know the coefficients $\alpha_m$
and $\beta_m.$ They will be expressed in section \ref{sect:anc} in
terms of the ground state averages of the conserved currents of the
$q$-boson lattice model.

\subsection{Bosonization of the $q$-boson lattice model: averages of the Noether currents \label{sect:anc}}

An important part of the bosonization procedure is the calculation
of the non-universal $c$-number parameters of the effective theory
(Luttinger parameter $K,$ sound velocity $v,$ coefficients
$\alpha_m$ and $\beta_m$) by establishing their correspondence with
the properties of the microscopic theory under consideration. This
is done for the $q$-boson lattice model in the present section. The
methodology used is standard for the Bethe-ansatz solvable models
\cite{H2-81}, so the presentation will be rather brief.

From the representation \eqref{jjel} it is evident that the
coefficients $\alpha_m$ and $\beta_m$ describe the response of the
system to large scale (smooth) variations of the local density
$\mathbf{j}^{(0)}_\tau$ and the particle current
$\mathbf{j}^{(0)}_x$. To calculate this response, we take our {\it
microscopic} model and calculate the variation of
$\mathcal{J}_\tau^{(m)}$ in response to the variation of
$\mathcal{J}_\tau^{(0)}$ and $\mathcal{J}_x^{(0)}$ in the vicinity
of the ground state. It follows from Eqs.~\eqref{Jmgs} and
\eqref{Jx0Jtau} that
\begin{equation}
\langle \mathcal{J}_x^{(0)}\rangle=0 \label{Jxgs}
\end{equation}
in the $q$-boson lattice model (recall that the symbol
$\langle\cdots\rangle$ denotes the ground state average). Consider a
homogeneous density variation, which satisfies Eq.~\eqref{Jxgs}.
Than the coefficient $\alpha_m$ in Eq.~\eqref{jjel} is defined by
the response of $\langle \mathcal{J}_\tau^{(m)} \rangle$ to this
density variation:
\begin{equation}
\alpha_m = \frac{\partial}{\partial D} \langle
\mathcal{J}_\tau^{(m)} \rangle, \label{alpha}
\end{equation}
where the density $D$ is given by Eq.~\eqref{rhodef}. To find
$\beta_m,$ we bosonize the operator $Q_\epsilon$ defined by
Eq.~\eqref{boost}:
\begin{equation}
\mathbf{Q}_\epsilon =\exp\left[i \epsilon \int dx\, \eta(x)
\mathbf{j}_\tau^{(0)}(x) \right] \label{Qebos}
\end{equation}
and apply to Eq.~\eqref{jjel} the same gauge transformation as was
discussed in section \ref{sect:lgtqblm}. We thus get
\begin{equation}
\left.\frac{d}{d\epsilon}\mathbf{j}^{(m)}_\tau\right|_{\epsilon=0} =
\left.\alpha_m
\frac{d}{d\epsilon}\mathbf{j}_\tau^{(0)}\right|_{\epsilon=0}+
\left.\beta_m
\frac{d}{d\epsilon}\mathbf{j}_x^{(0)}\right|_{\epsilon=0}
+\text{h.o.t.}, \qquad m=0,\pm1,\pm2,\dots.
\end{equation}
Assuming that $\eta(x)$ is a linear function of $x$,
\begin{equation}
\eta(x)=ax+b, \label{etaxlin}
\end{equation}
we use Eqs.~\eqref{boostinfc} and \eqref{boostel} with $a$ playing
the role of an infinitesimal variational parameter, and finally
obtain
\begin{equation}
\beta_m = - \frac{\delta^2 \chi^2 m}{2} \frac{\langle
\mathcal{J}_\tau^{(m)} \rangle}{\langle
\mathcal{J}_\tau^{(1)}\rangle}. \label{beta}
\end{equation}
Using Eq.~\eqref{Imtrans} one can rewrite the relations
\eqref{alpha} and \eqref{beta} in terms of the integrals of motion:
\begin{equation}
\alpha_m =\frac{\partial \langle I_m\rangle}{\partial \langle
I_0\rangle} \label{alphans}
\end{equation}
and
\begin{equation}
\beta_m = - \delta^2 \chi^2 \frac{m}2\frac{\langle
I_m\rangle}{\langle I_1\rangle}. \label{betans}
\end{equation}

Finally, we calculate the quantity\footnote{In the course of
calculating the function \eqref{g3def} the parameters $K$ and $v$
appear in the combination $Kv$ exclusively.} $Kv$ using the local
gauge transformation generated by the operator
$\mathbf{Q}_\epsilon,$ Eq.~\eqref{Qebos}. Comparing
Eqs.~\eqref{Pitauphi} and \eqref{jex} we find
\begin{equation}
\mathbf{j}_x^{(0)}(x) = -iKv\Pi(x)+\text{h.o.t.}
\label{jexcanonical}
\end{equation}
This gives the following commutation relation, with the use of
Eqs.~\eqref{jet}, \eqref{jex} and \eqref{canonical},
\begin{equation}
\left[\mathbf{j}_x^{(0)}(x), \mathbf{j}_\tau^{(0)}(y)\right]
=i\frac{Kv}{\pi} \frac{\partial}{\partial y} [\Pi(x), \phi(y)] = -
\frac{Kv}{\pi} \label{jjcomm}
\partial_x\delta(x-y).
\end{equation}
Now we apply the local gauge transformation generated by the
operator $\mathbf{Q}_\epsilon,$ Eq.~\eqref{Qebos}, to the
$x$-component of the current $\mathbf{j}^{(0)},$ Eq.~\eqref{jex}.
Using Eq.~\eqref{jjcomm} we get
\begin{multline}
\left.\frac{d}{d\epsilon}
\mathbf{j}_x^{(0)}(x,\epsilon)\right|_{\epsilon=0}=
\left.\frac{d}{d\epsilon}\left[\mathbf{Q}_\epsilon^{-1}
\mathbf{j}_x^{(0)}(x)\mathbf{Q}_\epsilon\right]\right|_{\epsilon=0}
\\= i \int dy\, \eta(y) \left[\mathbf{j}_x^{(0)}(x),
\mathbf{j}_\tau^{(0)}(y)\right] = -\frac{i K v}{\pi}\partial_x
\eta(x). \label{eqforKv}
\end{multline}
Imposing the condition \eqref{etaxlin} onto $\eta(x)$ and comparing
Eqs.~\eqref{eqforKv} and \eqref{boostel} we find
\begin{equation}
Kv= \frac{2\pi}{\delta^2 \chi^2} \frac{\langle
I_1\rangle}{M}=\frac{2\pi}{\delta \chi^2} \frac{\langle
I_1\rangle}{L}, \label{Kvans}
\end{equation}
where $L,$ $\delta,$ and $M$ are related by Eq.~\eqref{qCont}.

\section{Contour-independent integral identities \label{sect:stringid}}

The bosonization procedure, discussed in section
\ref{sect:bosonization}, is an effective tool for calculating the
correlation functions of the microscopic model in the low-energy
(long-distance) limit. Our task is, however, to calculate the {\it
local} function \eqref{g3def}. In the present section we construct
some contour-independent integral identities relating the short- and
long-distance correlation functions of the {\it microscopic} model
using its integrable structure. The long-distance contribution can
be found by making use of the bosonization procedure, and we are
thus getting ``for free'' an infinite set of non-trivial local
correlation functions. To calculate the function \eqref{g3def} we
need only one equation from this set; for other local observables
more equations could be needed. It will be also clear from the
calculations that the method proposed is general and can be used for
finding local correlation functions within other integrable models.

In section \ref{sect:gfcc} we calculate the coefficients $\alpha_0$
and $\beta_0,$ thus proving Eq.~\eqref{a1b0}. The method we use in
this proof is generalized in section \ref{sect:fbrim} to get some
nontrivial identities relating the conserved currents of the
microscopic ($q$-boson lattice) model with the properties of this
model in the bosonized limit. In section \ref{sect:cniir} we use
these identities to relate some local operator of the $q$-boson
lattice model to the properties of this model in the bosonized
limit. The result is given by Eq.~\eqref{stringanswer}. Its
continuum limit, studied in section \ref{sect:3bcf}, is the exact
expression \eqref{g3result} for the function $g_3(\gamma)$.

\subsection{Contour-independent integral identities and the bo\-so\-ni\-zed limit of the density operator \label{sect:gfcc}}

We consider in the present section a method of calculating the
coefficients $\alpha_0$ and $\beta_0.$ This method provides us with
a proof of Eq.~\eqref{a1b0} and, at the same time, it is an
important component of more general construction studied in section
\ref{sect:fbrim} and relating the short- and long-distance
correlation functions of the microscopic model.

Consider the commutator of the vertex operator $\mathcal{W}_m,$
Eq.~\eqref{vdef}, with the number operator $N.$ This commutator can
be calculated using formulas of section \ref{sect:fbtms} and gives
\begin{equation}
[N,\mathcal{W}_m(x)]=-m\mathcal{W}_m(x).
\end{equation}
The boson annihilation operator $\psi(x)$ of an arbitrary
microscopic theory, obeying the property
\begin{equation}
[N,\psi(x)]=-\psi(x), \label{Npsi}
\end{equation}
can, therefore, be written in the low-energy sector as follows
\begin{equation}
\psi(x)=c\mathcal{W}_1(x)+\text{h.o.t.} \label{psi->vert}
\end{equation}
Here $c$ is an unknown constant which depends on the structure of
the microscopic theory.

Using Eqs.~\eqref{Ndef}, \eqref{Nloc}, and \eqref{nocur}, we
represent Eq.~\eqref{Npsi} as follows:
\begin{equation}
\int_0^L dx\,\left[j^{(0)}_\tau(x),\psi(y)\right]=-\psi(y).
\label{jpsicomm}
\end{equation}
It would be tempting to substitute the bosonized expressions
\eqref{jmtau} and \eqref{psi->vert} into Eq.~\eqref{jpsicomm} and in
such a way obtain an equation for $\alpha_0$ and $\beta_0.$ One
should keep in mind, however, that the bosonization technique is an
approximation working in the long-distance limit. Whenever the
arguments of two operators, $j_\tau^{(0)}(x)$ and $\psi(y),$ are
close to each other, their product $j_\tau^{(0)}(x)\psi(y)$ cannot
be bosonized by simply bosonizing each of the two operators. To
circumvent this difficulty, we use a trick which will play a crucial
role for our studies of the function \eqref{g3def}. Let us explain
this trick in detail.

To begin with, we recall first the definition of a Green's formula
in the classical analysis. Consider a domain $\mathcal{D}$ in the
$(x,\tau)$ plane (we assume, for simplicity, that $\mathcal{D}$ is
compact and has a piecewise-smooth boundary $\partial \mathcal{D}$).
For any functions $P(x,\tau)$ and $Q(x,\tau)$ with continuous first
derivatives in $\mathcal{D},$ the following formula (Green's
formula) is valid
\begin{equation}
\iint\limits_\mathcal{D} dx\,d\tau\left(\frac{\partial Q}{\partial
x}- \frac{\partial P}{\partial \tau}\right)= \int\limits_{\partial
\mathcal{D}} P\,dx+Q\,d\tau. \label{Green}
\end{equation}
The boundary $\partial \mathcal{D}$ is oriented counterclockwise:
when going along $\partial \mathcal{D}$ the exterior of
$\mathcal{D}$ is kept on the right. We now consider the contour
integral
\begin{equation}
\int\nolimits_{\Gamma} P\,dx+Q\,d\tau, \label{Gc}
\end{equation}
where the contour $\Gamma$ is not necessarily closed. If
\begin{equation}
\frac{\partial Q}{\partial x}=\frac{\partial P}{\partial \tau}
\label{CR}
\end{equation}
in some region $\mathcal{D}^\prime$, than, according to
Eq.~\eqref{Green}, any deformation of $\Gamma$ within
$\mathcal{D}^\prime$ does not change the value of the integral
\eqref{Gc}.

\begin{figure}
\begin{center}
\includegraphics[clip,width=14cm]{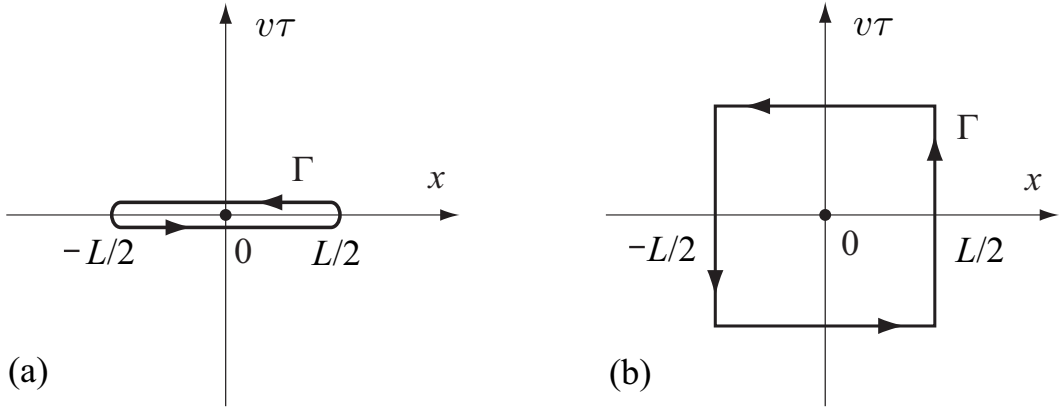}
\end{center}
\caption{In (a) we show the integration contour $\Gamma$ of
Eq.~\eqref{jpcont}. It consists of two horizontal segments of length
$L$, infinitesimally close to the $x$-axis. The integration contour
$\Gamma$ of Eq.~\eqref{jpc2} is shown in (b). It has the shape of a
square with side-length $L.$ Both contours are oriented
counterclockwise.} \label{fig:rhobos}
\end{figure}

As our next step, we perform a sequence of transformations of
Eq.~\eqref{jpsicomm}. It is clear that one can set $y=0$ and
integrate over the interval $(-L/2,L/2)$ without loss of generality.
Then we write
\begin{multline}
\int_{-L/2}^{L/2} dx\,\left[j^{(0)}_\tau(x),\psi(0)\right]=
\int_{-L/2}^{L/2} dx\, \left[
j^{(0)}_\tau(x,\tau\to+0)\psi(0,0)\right.\\\left.-\psi(0,0)j^{(0)}_\tau(x,\tau\to-0)
\right]= -\int_{\Gamma}dx\, Tj_\tau^{(0)}(x,\tau)\psi(0,0).
\label{jpcont}
\end{multline}
The second argument of the operators $j^{(0)}_\tau$ and $\psi$ is
the imaginary time. The symbol $T$ denotes imaginary time ordering,
defined by Eq.~\eqref{Tord}. The contour $\Gamma$ is shown in
Fig.~\ref{fig:rhobos}(a).

We now prove that the expression on the right hand side of
Eq.~\eqref{jpcont} is contour-independent. Consider an arbitrary
contour enclosing the point $(0,0)$ in the $(x,v\tau)$ plane. To
make contact with Eqs.~\eqref{Green}--\eqref{CR}, we write
\begin{equation}
P=-Tj_\tau^{(0)}(x,\tau)\psi(0,0), \qquad
Q=Tj_x^{(0)}(x,\tau)\psi(0,0),
\end{equation}
resulting in
\begin{equation}
\frac{\partial Q}{\partial x}-\frac{\partial P}{\partial\tau}=
\partial^\mu Tj_\mu^{(0)}(x,\tau) \psi(0,0). \label{qp1}
\end{equation}
Recall that $j_x^{(0)}$ is given by Eqs.~\eqref{J0xco1} and
\eqref{nocur}. The $T$-operator in the expression \eqref{qp1}
commutes with $\partial^\mu$ at all points in the $(x,v\tau)$ plane
except at the origin of the coordinate system. Therefore
\begin{equation}
\frac{\partial Q}{\partial x}-\frac{\partial P}{\partial\tau}=  T
\partial^\mu j_\mu^{(0)}(x,\tau) \psi(0,0)=0, \qquad
(x,v\tau)\ne(0,0),
\end{equation}
where the last equality is ensured by the continuity equation
\eqref{Noether1}. We have thus shown that the expression
\begin{multline}
\int_\Gamma Pdx+Qd\tau\equiv \int_\Gamma
-Tj_\tau^{(0)}(x,\tau)\psi(0,0)dx+Tj_x^{(0)}(x,\tau)\psi(0,0)d\tau
\\=\int_\Gamma dx_\mu \epsilon^{\mu\nu}
Tj_\nu^{(0)}(x,\tau)\psi(0,0), \label{pq2}
\end{multline}
where
\begin{equation}
\epsilon_{xx}=\epsilon_{\tau\tau}=0, \qquad \epsilon_{\tau x}=-1,
\qquad \epsilon_{x\tau}=1, \label{epmunu}
\end{equation}
is contour-independent, unless the contour crosses the origin of the
coordinate system. For the contour $\Gamma$ plotted in
Fig.~\ref{fig:rhobos}(a), the equation \eqref{pq2} reduces to
Eq.~\eqref{jpcont}. We have thus proved that the right hand side of
Eq.~\eqref{jpcont} is contour-independent, and Eq.~\eqref{jpsicomm}
can be written as follows:
\begin{equation}
\int_\Gamma dx_\mu \epsilon^{\mu\nu}
Tj_\nu^{(0)}(x,\tau)\psi(0,0)=-\psi(0,0). \label{jpc2}
\end{equation}

Now we choose the shape of the contour $\Gamma$ in Eq.~\eqref{jpc2}
as shown in Fig.~\ref{fig:rhobos}(b): a square with  side-length
$L$. For large system size, $L,$ the operators $j_\nu^{(0)}(x,\tau)$
and $\psi(0,0)$ are well-separated, and their product can be
bosonized using Eqs.~\eqref{jmtau}, \eqref{jmx} and
\eqref{psi->vert}. Equation \eqref{jpc2} then reduces to a condition
\begin{equation}
\alpha_0=1,
\end{equation}
as already announced in Eq.~\eqref{a1b0}.

The other statement announced in Eq.~\eqref{a1b0}, $\beta_0=0,$ can
be proven by considering the commutation relation of $\psi(x)$ with
the momentum operator $P,$ Eq.~\eqref{Pdef}:
\begin{equation}
[P,\psi(x)]=i\partial_x\psi(x). \label{Ppsicomm}
\end{equation}
We write, using Eqs.~\eqref{J0xco1} and \eqref{nocur},
\begin{equation}
[P,\psi(0)]=\frac{i}2\int_{-L/2}^{L/2}
\left[j_x^{(0)}(x),\psi(0)\right]=-\frac{i}2\int_{\Gamma}dx\,
Tj_x^{(0)}(x,\tau)\psi(0,0),
\end{equation}
where the contour $\Gamma$ is chosen as shown in
Fig.~\ref{fig:rhobos}(a). Like Eq.~\eqref{jpcont}, this expression
can be written in a contour-independent form, after which the
contour can be deformed to the shape shown in
Fig.~\ref{fig:rhobos}(b), and the operators $j^{(0)}$ and $\psi$ can
be bosonized. One gets for $[P,\psi(0)]$ in the bosonized limit
\begin{equation}
[P,\psi(0)]\to \frac{i}2\beta_0 c\mathcal{W}_1(0)+\text{h.o.t}
\end{equation}
On the other hand, it follows from Eq.~\eqref{psi->vert} that the
operator $\partial_x\psi(x)$ in the bosonized limit is proportional
to the operator $\partial_x\theta\mathcal{W}_1(x),$ whose anomalous
dimension is higher than that of $\mathcal{W}_1(x).$ Thus the only
way to satisfy Eq.~\eqref{Ppsicomm} is to require
\begin{equation}
\beta_0=0.
\end{equation}
This completes the proof of Eq.~\eqref{a1b0}. It is important to
stress that the result \eqref{a1b0} is valid for any interacting
system, the only condition that should be fulfilled is the existence
of well-defined number and momentum operators for the system.

\subsection{Contour-independent integral identities and the bo\-so\-ni\-zed limit of the Noether currents \label{sect:fbrim}}

We have considered in section \ref{sect:gfcc} a contour-invariant
integral representation of the operators $[N,\psi(x)]$ and
$[P,\psi(x)].$ This method establishes a connection between the
short- and long-distance properties of the microscopic theory. The
long-distance sector of the theory can be bosonized, with the result
of getting explicit answers for the correlation functions. We will
continue to work in the present section with the contour-invariant
integral representation, studying the Noether currents in the
Lieb-Liniger model. Recall that we do not distinguish the
Lieb-Liniger and $q$-boson lattice model unless the lattice
regularization is required explicitly. Therefore, to shorten
notation, we will use in the most cases the continuum space variable
$x$ instead of the discrete variable $n.$ The modifications
necessary to take into account the discreetness of the space
variable are obvious.

We introduce an operator
\begin{equation}
W^{(m)}(x,\tau)=\int_{-L/2}^{x} dx' \, j^{(m)}_\tau(x', \tau),
\label{odef}
\end{equation}
where $j^{(m)}_\tau$ is the $\tau$-component of the conserved
current $j^{(m)},$ Eq.~\eqref{nocur}. The object which will play a
crucial role in our further calculations is
\begin{multline}
G_\Gamma^{(n,m)}\equiv\int_\Gamma dx\, \left\langle
Tj_\tau^{(n)}(x,\tau)W^{(m)}(0,0)\right\rangle- d\tau\, \left\langle
Tj_x^{(n)}(x,\tau)W^{(m)}(0,0)\right\rangle\\=\int_\Gamma
dx_\mu\epsilon^{\mu\nu}\, \left\langle
Tj_\nu^{(n)}(x,\tau)W^{(m)}(0,0)\right\rangle, \label{Gidepsilon}
\end{multline}
where $\epsilon_{\mu\nu}$ is defined by Eq.~\eqref{epmunu}, and $T$
by Eq.~\eqref{Tord}. Like in section \ref{sect:gfcc}, denote the
terms in the integrand of Eq.~\eqref{Gidepsilon} as follows:
\begin{equation}
P=-\left\langle Tj_\tau^{(n)}(x,\tau)W^{(m)}(0,0)\right\rangle,
\qquad Q=\left\langle Tj_x^{(n)}(x,\tau)W^{(m)}(0,0)\right\rangle,
\label{PQ}
\end{equation}
therefore,
\begin{equation}
\frac{\partial Q}{\partial x}-\frac{\partial P}{\partial\tau}=
\left\langle \partial^\mu T j_\mu^{(n)}(x,\tau)
W^{(m)}(0,0)\right\rangle. \label{QxPt1}
\end{equation}
The $T$-operator in this expression commutes with $\partial_t$ at
all the points of $(x,v\tau)$ plane except the segment
\begin{equation}
\Upsilon=(-L/2\le x\le0,0),
\end{equation}
therefore
\begin{equation}
\frac{\partial Q}{\partial x}-\frac{\partial P}{\partial\tau}=
\left\langle T
\partial^\mu j_\mu^{(n)}(x,\tau) W^{(m)}(0,0)\right\rangle=0,
\qquad (x,\tau)\notin\Upsilon, \label{QxPt}
\end{equation}
where the last equality is ensured by the continuity equation
\eqref{Noether1}.

\begin{figure}
\begin{center}
\includegraphics[clip,width=14cm]{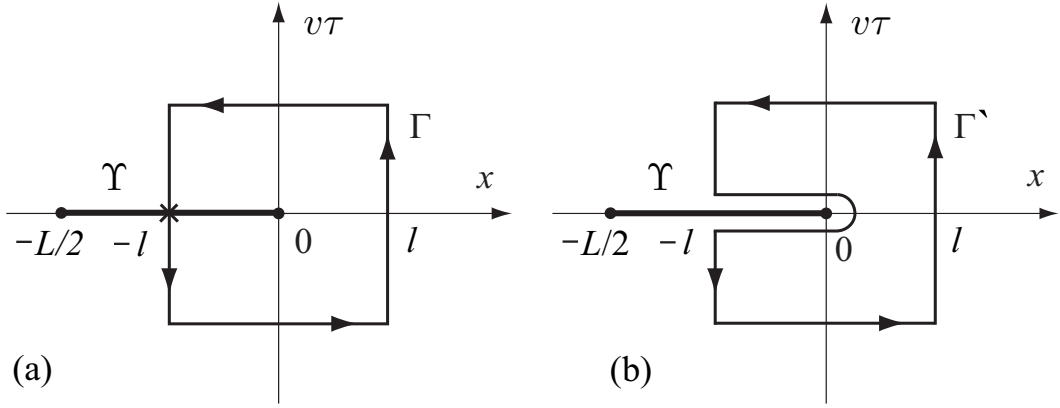}
\end{center}
\caption{In (a) we show the integration contour $\Gamma$ used in
section \ref{sect:fbrim}. It has the shape of a square with
side-length $L.$ The integration contour $\Gamma^\prime$ used in
section \ref{sect:cniir} is exhibited in (b).}  \label{fig:W}
\end{figure}

Let the contour $\Gamma$ in Eq.~\eqref{Gidepsilon} be a square
centered around the origin and with the side of length $l$, as shown
in Fig.~\ref{fig:W}a.
The segment $\Upsilon$ of the $(x,\tau)$ plane is shown there by a
thick solid line. The cross in Fig.~\ref{fig:W}a indicates the
intersection point of $\Gamma$ and $\Upsilon.$ Any deformation of
$\Gamma$ not changing the position of the intersection point, leaves
the value of $G_\Gamma^{(n,m)}$ unchanged.


We want to replace the exact currents $j_\mu^{(m)}$ entering
Eq.~\eqref{Gidepsilon} by their approximate expressions
\eqref{jmtau} and \eqref{jmx} obtained within the free boson theory.
There is, however, a problem: the operator $j_\nu^{(n)}(x,\tau)$ and
the operator $j_\tau^{(m)}(x^\prime,0)$ entering $W^{(m)}(0,0)$ are
not separated by the asymptotically large space-time interval in a
vicinity of the intersection point of $\Gamma$ and $\Upsilon$, and
one cannot apply Eqs.~\eqref{jmtau} and \eqref{jmx} to the operator
$j_\nu^{(n)}(x,\tau)j_\tau^{(m)}(x^\prime,0).$ To circumvent this
problem we use the following trick: Introduce an operator
\begin{equation}
\tilde W^{(m)}(x,\tau)\equiv -\int_{x}^{L/2}dx'\, j_\tau^{(m)}
(x',\tau). \label{odef1}
\end{equation}
Combining Eqs.~\eqref{odef} and \eqref{odef1}, one gets
\begin{equation}
W^{(m)}(x,\tau)-\tilde W^{(m)}(x,\tau)= \int_{-L/2}^{L/2}dx'\,
j_\tau^{(m)} (x',\tau)= I_m-\langle I_m \rangle,
\end{equation}
where $I_m$ are the integrals of motion of the $q$-boson lattice
model, discussed in section~\ref{sect:imqbm}. Then, we split the
contour $\Gamma$ into two parts:
\begin{equation}
\Gamma=\left\{\begin{array}{cc}\Gamma_{+}&x>0\\
\Gamma_{-}&x<0\end{array} \right. .  \label{Gsplit}
\end{equation}
Using Eqs.~\eqref{odef1}--\eqref{Gsplit} we rewrite the expression
\eqref{Gidepsilon} in the following form
\begin{multline}
G_{\Gamma}^{(n,m)}=\int_{\Gamma_+} dx_\mu \epsilon^{\mu \nu}
\left\langle T j_\nu^{(n)}(x,\tau) W^{(m)}(0,0) \right\rangle\\
+\int_{\Gamma_-} dx_\mu \epsilon^{\mu \nu} \left\langle T
j_\nu^{(n)}(x,\tau) \tilde W^{(m)}(0,0) \right\rangle+
\int_{\Gamma_-} dx_\mu \epsilon^{\mu \nu} \left\langle T
j_\nu^{(n)}(x,\tau) (I_m-\langle I_m \rangle ) \right\rangle.
\label{GGs}
\end{multline}
The third term on the right hand side of Eq.~\eqref{GGs} vanishes
because the ground state $|\mathrm{gs}\rangle$ is an eigenfunction
of $I_m.$ In the first (second) term the distance between the local
field $j_\nu^{(n)}(x,\tau)$ and the local field
$j_\tau^{(m)}(x^\prime,0)$ entering the operator $W^{(m)}(0,0)$ (the
operator $\tilde W^{(m)}(0,0)$) is larger than $l.$

A nontrivial output from Eq.~\eqref{GGs} comes in the limit
\begin{equation}
L\to\infty, \qquad l\to\infty, \qquad \frac{l}{L}\to 0.
\label{LRlim}
\end{equation}
Since the operators $j_\nu^{(n)}(x,\tau)$ and
$j_\tau^{(m)}(x^\prime,0)$ entering Eq.~\eqref{GGs} are separated by
a distance larger than $l,$ one can use the bosonized expressions
\eqref{jmtau} and \eqref{jmx} in the limit \eqref{LRlim}. Taking
into account Eq.~\eqref{Gbose}, one gets after some algebra
\begin{equation}
G_\Gamma^{(n,m)}= -\frac{Kv}{\pi} (\alpha_n \beta_m +\alpha_m
\beta_n). \label{identity}
\end{equation}

\subsection{Contour-independent integral identities: the main result \label{sect:cniir}}

In present section we relate the ground-state average of some
nontrivial {\it local} operator of the $q$-boson lattice model to
the properties of this model in the bosonized limit. Establishing
this relation, Eq.~\eqref{stringanswer}, we get all necessary
information for the calculation of the function $g_3(\gamma),$
Eq.~\eqref{g3def}.

We use the techniques developed in sections \ref{sect:gfcc} and
\ref{sect:fbrim}. Choose the contour $\Gamma^\prime$ in $(x,v\tau)$
plane as shown in Fig.~\ref{fig:W}(b). Namely, $\Gamma^\prime$
consists of the two horizontal segments, $(-l\le x\le \epsilon,+0)$
and $(-l\le x\le \epsilon,-0)$, and the square with the side $l,$
centered around the origin. The latter part is just the contour
$\Gamma$ used in sections \ref{sect:gfcc} and \ref{sect:fbrim}.
Using the techniques developed there we write
\begin{equation}
G_\Gamma^{(n,m)} +\int_{-l}^{\epsilon} dx \left\langle
\left[j^{(n)}_\tau (x,0), W^{(m)}(0,0)\right] \right\rangle =0.
\label{intmd}
\end{equation}
The parameter $\epsilon>0$ plays the role of a regularization
parameter, as it is necessary to treat unambiguously the behavior of
the integral in the vicinity of $x=0.$ We shall work with
Eq.~\eqref{intmd} in the limit \eqref{LRlim}, and we can therefore
use Eq.~\eqref{identity} to get
\begin{equation}
\int_{-l}^{\epsilon} dx \int_{-L/2}^0 d x'\, \left\langle
\left[j_\tau^{(n)}(x), j_\tau^{(m)}(x') \right] \right\rangle =
\frac{K v}{\pi}(\alpha_m \beta_n+\alpha_n\beta_m). \label{identity1}
\end{equation}

We remind the reader that we write the lattice regularization
explicitly only when it is necessary. Here is the proper place to do
it. Equation \eqref{identity1} then takes the form
\begin{equation}
\delta^2 \sum_{k=-l\delta^{-1}}^{\epsilon\delta^{-1}}
\sum_{p=-L\delta^{-1}/2}^0 \left\langle \left[j_\tau^{(n)}(k),
j_\tau^{(m)}(p) \right] \right\rangle = \frac{K v}{\pi}(\alpha_m
\beta_n+\alpha_n\beta_m). \label{identity2}
\end{equation}
Let us take the limit \eqref{LRlim}, keeping $\delta^{-1}$ finite.
Therefore
\begin{equation}
L\delta^{-1}\to\infty,\qquad l\delta^{-1}\to\infty \label{cond1}
\end{equation}
in Eq.~\eqref{identity2}. We then choose the parameter $\epsilon$
such that
\begin{equation}
\epsilon\delta^{-1}>m. \label{cond2}
\end{equation}

We set $n=-1$ and $m=2$ in Eq.~\eqref{identity2}. The local
operators $j_\tau^{(-1)}(k)$ and $j_\tau^{(2)}(p)$ are defined by
Eqs.~\eqref{J1}, \eqref{J2}, \eqref{J-m}, and \eqref{nocur}. Using
these definitions, we calculate the expression on the left hand side
of Eq.~\eqref{identity2} under the conditions \eqref{cond1} and
\eqref{cond2}:
\begin{multline}
\frac{\delta^2}{\chi^2} \left(1-\frac{\chi^2}{2} \right)^{-1}
\sum_{k=-l\delta^{-1}}^{\epsilon\delta^{-1}}
\sum_{p=-L\delta^{-1}/2}^0 \left \langle \left[j_\tau^{(-1)}(k),
j_\tau^{(2)}(p) \right] \right \rangle \\= \left\langle -\chi^2
B^{\dagger}_j B_{j+1}+\chi^4 B^\dagger_j B^\dagger_j B_j B_{j+1}
\right\rangle.
\end{multline}
Comparing this result with the right hand side of
Eq.~\eqref{identity2} one gets
\begin{equation}
\langle -\chi^2 B^{\dagger}_j B_{j+1}+\chi^4 B^\dagger_j B^\dagger_j
B_j B_{j+1} \rangle =\frac{1}{\chi^2} \left( 1-\frac{\chi^2}{2}
\right)^{-1} \frac{K v}{\pi}(\alpha_{-1}
\beta_2+\alpha_2\beta_{-1}). \label{strid}
\end{equation}
This expression relates the ground-state average of the local
operator of the microscopic model to the properties of this model in
the bosonized limit.

Let us combine Eqs.~\eqref{strid}, \eqref{J1}, and \eqref{Imtrans}
\begin{equation}
\langle B^\dagger_j B^\dagger_j B_j B_{j+1} \rangle =
\frac\delta{\chi^4} \frac{\langle I_1\rangle}{L} +\frac{1}{\chi^6}
\left( 1-\frac{\chi^2}{2} \right)^{-1} \frac{K v}{\pi}(\alpha_{-1}
\beta_2+\alpha_2\beta_{-1}), \label{stringans}
\end{equation}
and substitute into Eq.~\eqref{stringans} the results
\eqref{alphans}, \eqref{betans} and \eqref{Kvans}. This gives
\begin{equation}
\langle B^\dagger_j B^\dagger_j B_j B_{j+1} \rangle =
\frac\delta{\chi^4} \frac{\langle I_1\rangle}{L} \left[ 1+
\left(1-\frac{\chi^2}2\right)^{-1} \frac1{\chi^2}
\left(\frac{\partial\langle I_2\rangle}{\partial \langle
I_0\rangle}- 2\frac{\langle I_2\rangle}{\langle
I_1\rangle}\frac{\partial\langle I_1\rangle}{\partial \langle
I_0\rangle}\right) \right]. \label{stringanswer}
\end{equation}

\section{Three-body local correlation function \label{sect:3bcf}}

Working with the $q$-boson lattice model, we obtained in previous
sections all the formulas necessary to calculate $g_3(\gamma).$ The
remaining task is to take the continuum limit in these formulas and
to collect all them together. The limit is taken in section
\ref{sect:tqblmcl} and the formulas are collected together in
section \ref{sect:result}.

\subsection{Properties of $\langle I_m\rangle$ close to the continuum limit \label{sect:tqblmcl}}

In this section we continue our studies of $\langle I_m\rangle$ in
the $q$-boson lattice model, started in section \ref{sect:tqbm}. The
continuum limit \eqref{qCont} of the $q$-boson lattice model gives
the Lieb-Liniger model, Eq.~\eqref{qContHam}.  The limit in which we
are interested is the continuum limit together with the extra
condition $L\to\infty,$ coming from Eq.~\eqref{LRlim}. Therefore, we
consider
\begin{equation}
\delta\to0, \qquad M\to\infty, \qquad \kappa\to0, \qquad L\to\infty,
\label{tl1}
\end{equation}
where
\begin{equation}
L=\delta M, \qquad q=e^{\kappa}, \qquad c/2=\kappa\delta^{-1}.
\label{tl2}
\end{equation}

Let us renormalize the quasi-momenta $p_j$ used in section
\ref{sect:tqbm}:
\begin{equation}
k_j=\delta^{-1} p_j.
\end{equation}
The kernel \eqref{qkernel} can be represented in the limit
\eqref{tl1} as follows
\begin{equation}
K(p)=\delta^{-1}K(k), \qquad K(k)= \frac{2 c}{c^2+k^2}+
\frac{\delta^2}{6}c + \frac{ \delta^4}{360}\left(3
ck^2-c^3\right)+\dots . \label{Kexp}
\end{equation}
When written in terms of the variables $k_j,$ the normalization
condition \eqref{qnorm} becomes
\begin{equation}
n= \int_{-\Lambda(\delta)}^{\Lambda(\delta)} dk\, \rho(k,\delta),
\label{nTBA}
\end{equation}
and the Lieb-Liniger equation \eqref{Lieb} becomes
\begin{equation}
\rho(k,\delta)-\frac{1}{2\pi}\int_{-\Lambda(\delta)}^{\Lambda(\delta)}d\tilde
k\, K(k-\tilde k) \rho(\tilde k,\delta)= \frac{1}{2\pi}.
\label{rhoTBA}
\end{equation}
Here the ground-state density $D$ is defined by Eq.~\eqref{rhodef}.
Note that the functions $\Lambda$ and $\rho$ in Eqs.~\eqref{nTBA}
and \eqref{rhoTBA} are different from those used in
Eqs.~\eqref{qnorm} and \eqref{qkernel}. We, however, use the same
symbols for these two couples, since we shall work with
Eqs.~\eqref{nTBA} and \eqref{rhoTBA} exclusively.

The ground state expectation values of $I_1$ and $I_2,$
Eqs.~\eqref{I1g} and \eqref{I2g}, can be represented in the limit
\eqref{tl1} as follows
\begin{equation}
\frac{\langle I_1\rangle}L = \chi^2 \left(D- \frac{\delta^2}{2!} h_2
+\frac{\delta^4}{4!} h_4+ \dots \right), \label{I1exp}
\end{equation}
and
\begin{equation}
\frac{\langle I_2\rangle}L = \chi^2 \left(1-\frac{\chi^2}{2}\right)
\left(D-\frac{(2\delta)^2}{2!} h_2 +\frac{(2\delta)^4}{4!} h_4+
\dots \right), \label{I2exp}
\end{equation}
where
\begin{equation}
h_m(\delta)=\int_{-\Lambda(\delta)}^{\Lambda(\delta)}dk\,
\rho(k,\delta)k^m. \label{hdef}
\end{equation}
Equations \eqref{I1exp} and \eqref{I2exp} are not series expansions
in powers of $\delta$ since the functions $h_m$ themselves depend on
$\delta$ through the quasi-momentum distribution $\rho$ and through
$\Lambda.$ The series expansion of $h_m$ in powers of $\delta$ is
\begin{equation}
h_m(\delta)=h_m^{(0)} + \delta^2 h_m^{(1)} + \dots \label{hmexp}
\end{equation}

Let us consider Eqs.~\eqref{nTBA} and \eqref{rhoTBA} at $\delta=0:$
\begin{equation}
D= \int_{-\Lambda(0)}^{\Lambda(0)} dk\, \rho(k,0) \label{nTBA0}
\end{equation}
and
\begin{equation}
\rho(k,0)-\frac{1}{2\pi}\int_{-\Lambda(0)}^{\Lambda(0)}d\tilde k\,
\frac{2 c}{c^2+(k-\tilde k)^2} \rho(\tilde k,0)= \frac{1}{2\pi}.
\label{rhoTBA0}
\end{equation}
Following Ref.~\cite{LL-63} we change the variables
\begin{equation}
k=\Lambda(0) z, \qquad c=\Lambda(0)\alpha, \qquad \rho(\Lambda(0)
z,0)=\sigma(z).
\end{equation}
When written in these variables, Eqs.~\eqref{nTBA0} and
\eqref{rhoTBA0} become Eqs.~\eqref{b2} and \eqref{b1}, respectively.
The function $h_m^{(0)}$ introduced by \eqref{hmexp} can be written
as
\begin{equation}
h_m^{(0)}(D, c)=D^{m+1} \epsilon_m\left(\gamma\right), \label{h20}
\end{equation}
where the function $\epsilon_m(\gamma),$ given by Eq.~\eqref{b3},
depends on the dimensionless parameter $\gamma,$ Eq.~\eqref{gamma},
only. One verifies readily that $h_m^{(0)}$ satisfies the following
differential equation
\begin{equation}
D\frac{\partial}{\partial D} h_m^{(0)}+ c\frac{\partial}{\partial c}
h_m^{(0)} - (m+1)h_m^{(0)}=0. \label{hom}
\end{equation}

Next, we consider the solution of Eqs.~\eqref{nTBA} and
\eqref{rhoTBA} to second order in $\delta.$ Substituting
Eq.~\eqref{Kexp} into \eqref{rhoTBA} and keeping the terms up to the
order of $\delta^2$ we find
\begin{equation}
\rho(k,\delta)-\frac{1}{2\pi}\int_{-\Lambda(\delta)}^{\Lambda(\delta)}d\tilde
k \frac{2 c}{c^2+(k-\tilde k)^2} \rho(\tilde k,\delta)= \frac{1}{2
\pi } \left(1+\frac{c\delta^2 }{6}D\right). \label{TBA1ord}
\end{equation}
By rescaling the quasi-momentum distribution function
\begin{equation}
\rho(k,\delta) = \left(1+\frac{c\delta^2}{6}D\right)
\tilde\rho(k,\delta)
\end{equation}
we find that $\tilde\rho(k,\delta)$ satisfies the integral equation
\begin{equation}
\tilde\rho(k,\delta)-\frac{1}{2\pi}\int_{-\Lambda(\delta)}^{\Lambda(\delta)}d\tilde
k \frac{2 c}{c^2+(k-\tilde k)^2} \tilde\rho(\tilde k,\delta)=
\frac{1}{2\pi} \label{rhoTBAd}
\end{equation}
with the condition \eqref{nTBA} renormalized as follows:
\begin{equation}
D\left(1-\frac{c\delta^2}{6}D\right) =
\int_{-\Lambda(\delta)}^{\Lambda(\delta)}dk\, \tilde\rho(k,\delta).
\label{nTBAd}
\end{equation}
It is clear from comparison of Eqs.~\eqref{nTBAd} and
\eqref{rhoTBAd} with Eqs.~\eqref{nTBA0} and \eqref{rhoTBA0} that the
function $h_2,$ Eq.~\eqref{hdef}, can be represented to order
$\delta^2$ as follows
\begin{equation}
h_2(D,c) =\left(1+\frac{c\delta^2 D}{6}\right)
\int_{-\Lambda(\delta)}^{\Lambda(\delta)}dk\,
k^2\tilde\rho(k,\delta)= \left(1+\frac{c\delta^2
D}{6}\right)h_2^{(0)}\left(D-\frac{c\delta^2 D^2}{6}, c\right).
\end{equation}
Expanding this equation and dropping terms of higher order than
$\delta^2,$ we get
\begin{equation}
h_2= h_2^{(0)}+ \frac{c\delta^2 D}{6} \left(h_2^{(0)}-
D\frac{\partial h_2^{(0)}}{\partial D}  \right). \label{hhom}
\end{equation}
Finally, combining Eqs.~\eqref{hom} and \eqref{hhom}, we get the
following differential equation for $h_2:$
\begin{equation}
D\frac{\partial h_2}{\partial D}  + c\frac{\partial h_2}{\partial c}
-3 h_2= \frac{c \delta^2 D}{3}\left(h_2- D\frac{\partial
h_2}{\partial D}  \right)+\dots \label{homcor}
\end{equation}
where we have dropped all the terms of the order higher than
$\delta^2.$

\subsection{Three-body local correlation function: the result \label{sect:result}}

We are now in the position to derive the main result of this paper:
equation \eqref{g3result} for the three-body local correlation
function $g_3(\gamma),$ Eq.~\eqref{g3def}, of the Lieb-Liniger model
in the limit \eqref{tl}. To do this, we take the limit \eqref{tl1}
in the ground-state averages of local operators of the $q$-boson
model, which were obtained in earlier sections. The corresponding
calculations are straightforward but rather lengthy, and we shall
only sketch their main steps.

The desired ground state expectation value \eqref{g3def} in the
Lieb-Liniger model can be obtained by taking the limit \eqref{tl1}
of the following $q$-boson local field:
\begin{multline}
\left\langle (1-\chi^2) B^\dagger_j B^\dagger_j B_j B_{j+1}+
\left(1-\frac{\chi^2}{2}\right) q\frac{d}{dq} (B_j^\dagger B_{j+1})
\right \rangle\\
=-\frac{c \delta^4}{3}\left[
\langle\psi^\dagger(x)^3\psi(x)^3\rangle+\dots\right], \label{Bpsi}
\end{multline}
where the dots denote terms of higher order in $\delta.$ To prove
Eq.~\eqref{Bpsi} we use Eq.~\eqref{HFTidentity} and then take the
limit \eqref{tl1} in the way discussed in section \ref{sect:hqblm},
Eqs.~\eqref{xqexp}--\eqref{xdn}. After rather lengthy algebra we got
the right hand side of Eq.~\eqref{Bpsi}.

As our next step we rewrite the left hand side of Eq.~\eqref{Bpsi}
using the identities \eqref{stringanswer} and \eqref{HFTlocal}
\begin{multline}
\left\langle (1-\chi^2) B^\dagger_j B^\dagger_j B_j B_{j+1}+
\left(1-\frac{\chi^2}{2}\right) q\frac{d}{dq} (B_j^\dagger B_{j+1})
\right \rangle= \left(1-\frac{\chi^2}{2}\right) q\frac{d}{d q}
\left(\frac{\langle I_1\rangle}{L}\frac\delta{\chi^2}\right)\\
+ (1-\chi^2)\frac\delta{\chi^4} \frac{\langle I_1\rangle}{L} \left[
1+ \left(1-\frac{\chi^2}2\right)^{-1} \frac1{\chi^2}
\left(\frac{\partial\langle I_2\rangle}{\partial \langle
I_0\rangle}- 2\frac{\langle I_2\rangle}{\langle
I_1\rangle}\frac{\partial\langle I_1\rangle}{\partial \langle
I_0\rangle}\right) \right] \label{Btp}
\end{multline}
and take the limit \eqref{tl1} in the resulting expression.
Substituting the expansions \eqref{I1exp} and \eqref{I2exp} into the
right hand side of Eq.~\eqref{Btp} one arrives at
\begin{multline}
\left\langle (1-\chi^2) B^\dagger_j B^\dagger_j B_j B_{j+1}+
\left(1-\frac{\chi^2}{2}\right) q\frac{d}{dq} (B_j^\dagger B_{j+1})
\right \rangle\\
=\left(\frac{\delta^3}{2}-\frac{\delta^2}{c}\right)
\left(D\frac{\partial h_2}{\partial D} + c \frac{\partial
h_2}{\partial c}- 3h_2 \right)- \delta^4 \left[ \frac{c}{12}
\left(D\frac{\partial h_2}{\partial D} + 3c\frac{\partial
h_2}{\partial c}  - 3 h_2\right)\right.\\
-\left. \frac{1}{12 c} \left(7D\frac{\partial  h_4}{\partial D}+ c
\frac{\partial h_4}{\partial c}  -15h_4 \right)+
\frac{h_2}{c}\frac{\partial h_2}{\partial D}  \right]. \label{Bth}
\end{multline}
where we have dropped terms of higher order than $\delta^4.$ Using
Eq.~\eqref{homcor} to transform the first term on the right hand
side of Eq.~\eqref{Bth} one can see that the expansion \eqref{Bth}
starts from the terms of order $\delta^4.$ Comparing these leading
order terms with the term written explicitly on the right hand side
of Eq.~\eqref{Bpsi} and replacing $h_m$ with $h_m^{(0)},$
Eq.~\eqref{h20}, one gets after some algebra the final result,
Eq.~\eqref{g3result}.

\subsection*{Acknowledgments}

The authors would like to thank N.M.~Bogoliubov and V.~Tarasov for
helpful discussions. M.B.~Zvonarev's work was supported by the
Danish Technical Research Council via the Framework Programme on
Superconductivity and by the Swiss National Fund for research under
MANEP and Division II.

\end{document}